\documentclass[12pt,preprint]{aastex}
\usepackage{graphicx}
\usepackage{longtable}
\shorttitle{}
\shortauthors{Carruba et al.}
\begin{document}

\title{Detection of the YORP Effect for Small Asteroids\\ in the Karin Cluster}
\author{V. Carruba$^{1,2}$, D. Nesvorn\'{y}$^{2}$, D. Vokrouhlick\'{y}$^{3}$}
\affil{(1) UNESP, Univ. Estadual Paulista, Grupo de Din\^{a}mica Orbital e
  Planetologia, Guaratinguet\'{a}, SP, 12516-410, Brazil}
\affil{(2) Department of Space Studies, Southwest Research Institute, Boulder, 
  CO, 80302, USA}
\affil{(3) Institute of Astronomy, Charles University, V
  Hole\v{s}ovi\v{c}k\'{a}ch 2, Prague, CZ-18000, Czech Republic}
\begin{abstract}
The Karin cluster is a young asteroid family thought to have formed only 
$\simeq 5.75$~My ago. The young age can be demonstrated by numerically 
integrating the orbits of Karin cluster members backward in time and showing 
the convergence of the perihelion and nodal longitudes (as well as other 
orbital elements). Previous work has pointed out that the convergence is not 
ideal if the backward integration only accounts for the gravitational 
perturbations from the Solar System planets. It improves when the 
thermal radiation force known as the Yarkovsky effect it is accounted for. 
This argument can be used to estimate the spin obliquities of the Karin cluster 
members. Here we take advantage of the fast growing membership of the 
Karin cluster and show that the obliquity distribution of diameter 
$D\simeq 1-2$ km Karin asteroids is bimodal, as expected if the YORP effect 
acted to move obliquities toward the extreme values ($0^\circ$ or $180^\circ$). 
The measured magnitude of the effect is consistent with the standard YORP model.
The surface thermal conductivity is inferred to be $0.07$-0.2 W m$^{-1}$ K$^{-1}$ 
(thermal inertia $\simeq 300-500$ J m$^{-2}$ K$^{-1}$s$^{-1/2}$). We find that the 
strength of the YORP effect is roughly $\simeq 0.7$ of the nominal strength 
obtained for a collection of random Gaussian spheroids. These results are 
consistent with a surface composed of rough, rocky regolith. The obliquity 
values predicted here for 480 members of the Karin cluster can be validated by 
the light-curve inversion method.  
\end{abstract}
%
%\keywords Minor planets, asteroids: general -- celestial mechanics  
%
%________________________________________________________________

\section{Introduction} \label{sec: intro}
The Karin family with the estimated age of $5.75\pm0.05$~My is one of 
the youngest families in the main belt \citep{Nesvorny_2002, Nesvorny_2004}.  
Because of its recent formation, it is possible to numerically integrate 
the orbits backward in time and demonstrate the young age by showing that 
the orbits of individual members converge together at the time 
of the parent body breakup. Improving on previous work, 
\citet{Nesvorny_2004} have shown that a precise reconstruction of the orbital 
histories requires that the Yarkovsky effect is taken into account in 
the backward integration. This allowed them to infer the semi-major axis 
drift rate for individual members of the Karin cluster and verify, for the 
first time, how the Yarkovsky effect operates on the main belt asteroids 
over million-year-long timescales. A by-product of this study was a 
determination of spin obliquities for 70 individual members of the Karin 
cluster with absolute magnitudes $H<16$ (roughly diameters $D>2$~km for albedo
$p_{\rm V}=0.2$).    

Many new asteroids have been discovered since the last dynamical analysis of 
the Karin cluster.  Here we repeat the analysis of \citet{Nesvorny_2004} with 
an orbital catalog that contains nearly seven times more asteroids 
than what available back in 2004. In Sect.~\ref{sec: fam_ide}, we revise the 
Karin family membership by applying the usual clustering method on the new 
orbital catalog. The taxonomical and albedo interlopers are eliminated.  
We then apply a more stringent criterion of the Karin family membership by 
requiring that orbits converge to each other $\simeq 5.75$~My ago.  In 
Sect.~\ref{sec: karin_drift}, we use the method developed in 
\citet{Nesvorny_2004} to estimate the Yarkovsky drift rates of individual 
bodies. This data is compared to the theoretical expectations for the 
Yarkovsky effect.

We find that the distribution of spin obliquities $\varepsilon$ of 
small Karin members ($D\simeq 1-2$~km) is bimodal with only very 
few values near $\varepsilon=90^\circ$ and peaks for smaller and larger 
obliquities (Sect.~4). It is shown that this obliquity distribution is 
consistent with an initially random orientation of spin axes that was 
modified by the YORP effect (Sects. 5-7; e.g. Rubincam 2000, 
\v{C}apek \& Vokrouhlick\'y 2004). In Sect.~\ref{sec: Karin_yorp}, we 
apply a standard YORP model to estimate the thermal conductivity and 
calibrate the strength of the YORP effect. The results are discussed 
in Sect.~9. Finally, we perform new numerical simulations with the 
Yarkovsky force and/or gravitational perturbations of (1) Ceres 
(Sect.~\ref{sec: yarko_num}), and discuss the latter as a stochastic factor 
that sets firm limits on what can be achieved with this type of study. 
Section \ref{sec: conc} presents our conclusions.

\section{Family identification} \label{sec: fam_ide}
To define the Karin cluster membership, we first turned our attention to 
the family identification data from \citet{Nesvorny_2015}.  In that work, 
the Karin cluster was identified using the Hierarchical Clustering Method 
(HCM hereafter), and a velocity cutoff of 10 m s$^{-1}$ in the domain of 
the proper orbital elements $(a,e,\sin i)$ \citep[see Table 2 in][for
further details]{Nesvorny_2015}. To eliminate possible 
interlopers, we adopted the classification scheme of \citet{DeMeo_2013}. 
Specifically, we used the fourth release of the Sloan Digital 
Sky Survey-Moving Object Catalog \citep[SDSS-MOC4;][]{Ivezic_2001}, 
and computed the $gri$ slope and $z' -i'$ colors. In addition, we 
used information from three major photometric/spectroscopic surveys: 
the Eight-Color Asteroid Survey \citep[ECAS;][]{Zellner_1985, Tholen_1989}, 
Small Main Belt Spectroscopic Survey
\citep[SMASS;][]{Xu_1995, Bus_2002a, Bus_2002b}, and Small Solar 
System Objects Spectroscopic Survey \citep[S3OS2;][]{Lazzaro_2004}. There were
13 objects with known taxonomical information in total, six of which have a 
C-complex taxonomy and are therefore incompatible with the S-type taxonomy 
of the Karin cluster.  After eliminating these objects we end up with a 
sample of 535 Karin family members.  

To account for possible members of the Karin cluster that may have been 
excluded by the velocity cutoff used in \citet{Nesvorny_2015}, we 
define a box in proper $(a,e,\sin i)$ space with the following ranges: 
$2.855$ to $2.878$~au in $a$, $0$ to $0.1$ in $e$, and $0.0122$ to $0.0611$
in $\sin i$. These values correspond to the full range of $(a,e,\sin i)$ values 
in the Karin cluster from \citet{Nesvorny_2015}, plus a margin of $0.002$~au 
in $a$, and $0.03$ in $e$ and $\sin i$. After eliminating SDSS-MOC4 
interlopers, we were left with a sample of $1117$ additional objects.  
Of these, only $8$ objects have known albedo values $p_{\rm V}<0.1$ 
\citep{Masiero_2012}, and can be potential albedo interlopers.

We proceed by computing the components $(v_r,v_t,v_W)$ the terminal 
ejection velocity $V_{\rm ej}$ from the Gauss equations 
\citep[e.g.,][]{Murray_1999}
\begin{eqnarray}
\frac{\delta a}{a} &=& \frac{2}{na\,(1-e^2)^{1/2}}[(1+e \cos f)\, \delta v_t +
 (e \sin f)\, \delta v_{r}]\; , \nonumber \\
\delta e &=& \frac{(1-e^2)^{1/2}}{na}\left[\frac{e+\cos f +e \cos^2 f}{1+e 
 \cos f } \,\delta v_t+\sin f\, \delta v_r\right]\; , \nonumber \\
\delta i &=& \frac{(1-e^2)^{1/2}}{na} \frac{\cos{(\omega+f)}}{1+e \cos f } 
 \, \delta v_W. \label{eq: gauss_3}
\end{eqnarray}
where $\delta a = a-a_{\rm ref}$, $\delta e = e-e_{\rm ref}$ and 
$\delta i= i-i_{\rm ref}$ with $a_{\rm ref}$, $e_{\rm ref}$ and $i_{\rm ref}$ 
being a reference value, and $f$ and $\omega$ are the true anomaly and 
perihelion argument of the disrupted parent body at the time of the breakup.  
Here we used $f=30^{\circ}$ and $f+\omega = 50.5^{\circ}$ \citep{Nesvorny_2004}. 

\begin{figure}
\epsscale{0.8}
\plotone{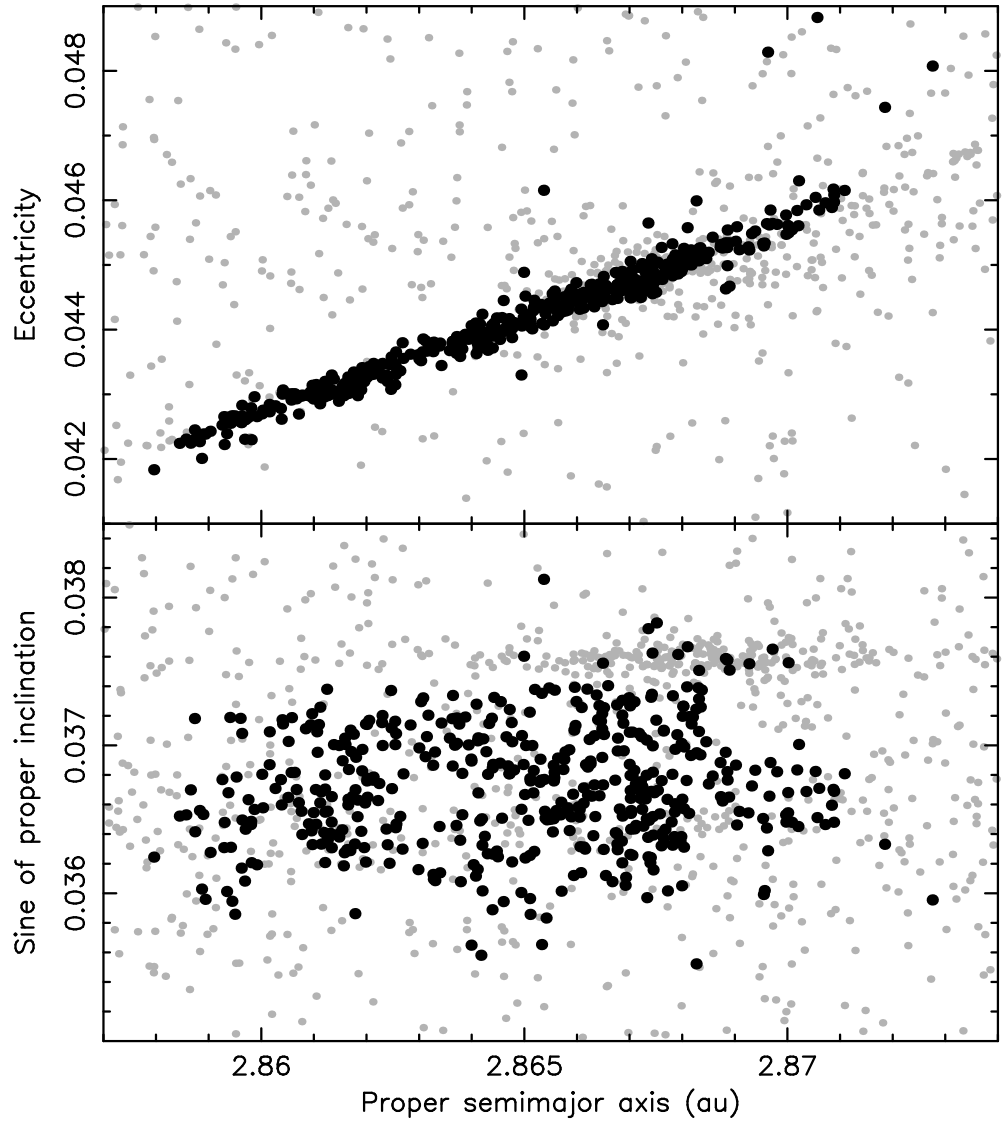}
\caption{A $(a,e)$ (top panel) and $(a,\sin i)$ (bottom panel) projection
 of members of the Karin cluster that satisfy the selection criteria
 discussed in Sect.~\ref{sec: karin_drift} (480 members, full black dots), 
 and of asteroids in the local background (full gray dots). The alignment
 of background gray objects seen for $\sin i \simeq 0.0375$ is the Koronis(2) 
 family \citep{Molnar_2009}.}
\label{fig: Karin_aei}
\end{figure}

We find that the HCM members of the Karin family have 
$V_{\rm ej} < 70$ m s$^{-1}$.  As a final membership filter we therefore 
include bodies in the extended set with $V_{\rm ej}<80$ m s$^{-1}$ (i.e., 
with a 10 m s$^{-1}$ buffer). In total, 489 asteroids in the Karin family 
and 189 in an extended family pass this filter.  A plot of the orbital 
distribution of 480 Karin family members, after applying additional criteria
discussed in the following text, is shown in Fig.~\ref{fig: Karin_aei}. 

To reconstruct the past orbital history of Karin cluster members, we 
numerically integrated the orbits of all $489+189=678$ potential members 
with the symplectic integrator known as $SWIFT\_MVSF$ \citep{Levison_1994}, 
modified by \citet{Broz_1999} to include the online filtering 
of the osculating elements.  The integration included the gravitational 
effects of all Solar System planets (the radiation forces were ignored).  
The initial velocity vectors of asteroids and planets were multiplied by 
$-1$ such that, effectively, the orbits are tracked back into the 
past.  The normal orbital longitudes $\Omega$ and $\varpi$ were
recovered from this simulation by using relationships
\begin{eqnarray}
 \Omega &=& \Omega^*+180^{\circ}\; , \nonumber \\
 \varpi &=& \Omega^* -\omega^*\; , \label{eq: time_rev}
\end{eqnarray} 
where $\Omega^*$ and $\omega^*$ are the nodal longitude and perihelion 
argument computed from the backward integration with $SWIFT\_MVSF$.
The integration time step was set to be 1 day.

\begin{figure}
\epsscale{1.0}
\plotone{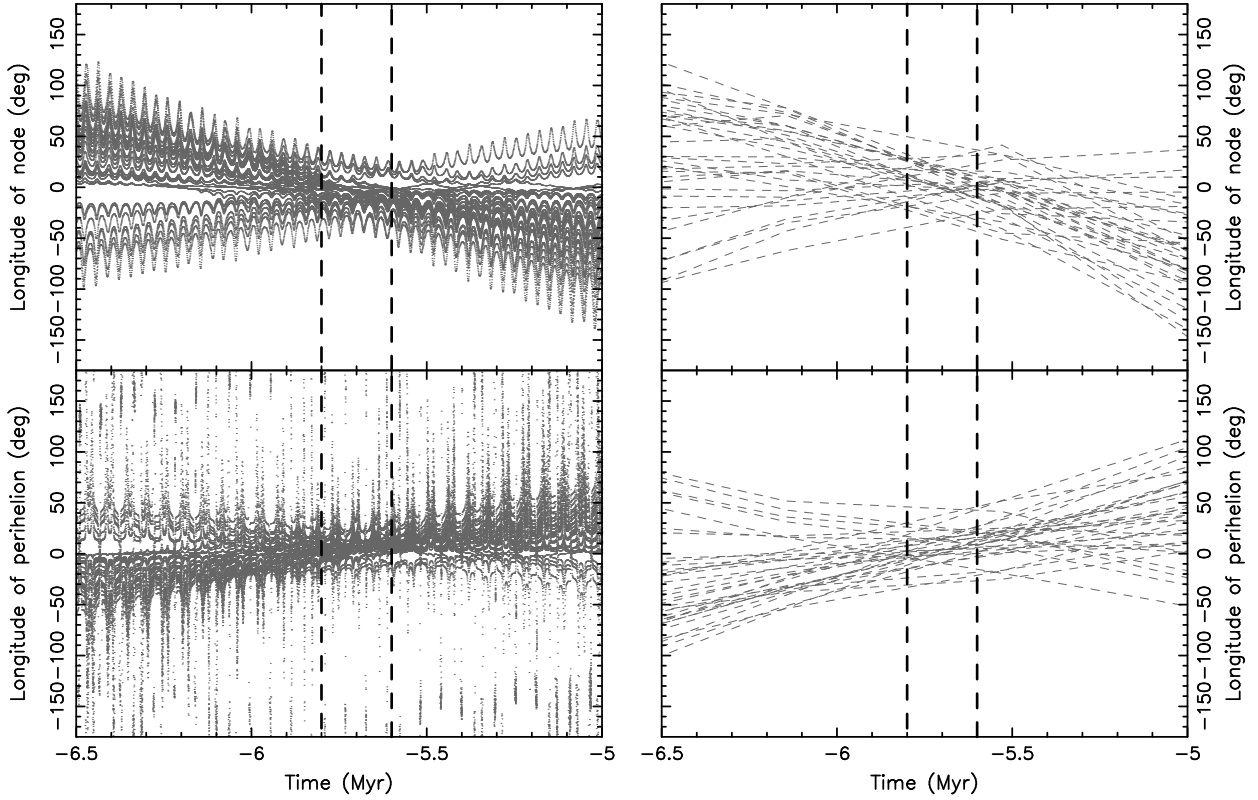}
\caption{Past evolution of the osculating (left panel) and mean
 (right panel) $\Omega$ and $\varpi$ angles for 34 large members of the
 Karin cluster. The vertical dashed lines delimit the time interval between 
 $-5.6$ and $-5.8$~My. The mean perihelion and nodal longitudes 
 were obtained using the Frequency Modified Fourier Transform (FMFT) method of 
 \citet{Sidli_1997}. The convergence of angles of all these large members 
 of the Karin cluster were originally reported in \citet{Nesvorny_2004}.} 
\label{fig: karin_angles}
\end{figure}

Figure~\ref{fig: karin_angles} shows the result of our backward simulation. 
We plot there $\Delta {\Omega} =  {\Omega}-{{\Omega}}_{\rm Karin}$ and 
$\Delta {\varpi} =  {\varpi}-{{\varpi}}_{\rm Karin}$, where $\Omega_{\rm Karin}$ 
and $\varpi_{\rm Karin}$ are the orbital longitudes of (832) Karin. Note that 
the angles converge in Fig.~\ref{fig: karin_angles} in the time interval 
between -5.6 and -5.8 My, which is a clear evidence that the Karin
cluster formed at that time (see also \citet{Novakovic_2012} for
details on the method of convergence of orbital angles as a membership 
criteria). From the 678 member candidates identified 
above we found that 576 objects have angles converging with 
$\Delta {\Omega} < 60^\circ$ and $\Delta {\varpi} < 60^\circ$ 
at $-5.8<t<-5.6$ My. These 576 objects represent our final
membership list. Relative to \citet{Nesvorny_2004} we identified 
479 new members of the Karin cluster.

\section{Measurement of the Yarkovsky Drift} \label{sec: karin_drift}
The convergence of angles in Fig.~\ref{fig: karin_angles} is not 
ideal because our numerical integration only accounted for the 
gravitational effects of planets and ignored all else. In reality, 
the orbits of small members of the Karin cluster are 
affected by the Yarkovsky effect that arises as a recoil force from a 
directional emission of the thermal radiation \citep[e.g.,][]{Bottke_2006}.
The main orbital effect 
of the Yarkovsky force is to either decrease or increase the semi-major 
axis of an orbit. Since the precession frequency of angles $\Omega$ and 
$\varpi$ depends on the semi-major axis, the Yarkovsky effect is thus 
expected to influence the convergence of $\Omega$ and $\varpi$.
This dependence can be used to determine the Yarkovsky drift rates 
for individual members of the Karin cluster \citep{Nesvorny_2004}.

According to \citet{Nesvorny_2004}, the values of $\Delta {\Omega}_{j}$ 
and $\Delta {\varpi}_{j}$ for asteroid $j$ at time $t=-(\tau+\Delta t)$ are:
\begin{equation}
 \Delta {\Omega}_{j}(t) = -\frac{1}{2}\frac{\partial s}{\partial {a}}(\delta a_{j}-
 \delta a_{1}) \tau -(s_j-s_1) \Delta t \; , \label{eq: Delta_Om}
\end{equation}
\begin{equation}
 \Delta {\varpi}_{j}(t) = -\frac{1}{2}\frac{\partial g}{\partial {a}}(\delta a_{j}-
 \delta a_{1}) \tau -(g_j-g_1) \Delta t\; , \label{eq: Delta_varpi}
\end{equation}
where $\tau$ is the estimated family age, $\Delta t$ is a small correction and
$\delta a_{j}$ is the total semi-major axis drift over time $\tau$. Here
we neglected the initial spread of these angles produced 
by $V_{\rm ej}$, which should be of the order of 1$^\circ$ \citep{Nesvorny_2004}. 
Index $j=1$ refers to (832) Karin. Quantities $\partial s/\partial {a}$ and 
$\partial g/\partial {a}$ define how the nodal and apsidal precession 
frequencies change with $a$. Here we adopt $\partial s/\partial {a}_P=-70.0$ 
arcsec yr$^{-1}$ au$^{-1}$ and $\partial g/\partial {a}_P=94.4$ arcsec yr$^{-1}$ 
au$^{-1}$ \citep{Nesvorny_2004}. Corrections $(s_j-s_1) \Delta t$ and 
$(g_j-g_1) \Delta t$ vanish when $\Delta t = 0$.  See \citet{Nesvorny_2004} 
for further discussion of  Eqs.~(\ref{eq: Delta_Om}) and 
(\ref{eq: Delta_varpi}). 

By solving these two equations we can obtain the values of $\Delta a_{j}
=\delta a_{j}-\delta a_{1}$ 
required to compensate for $\Delta {\Omega}$ and $\Delta {\varpi}$ obtained 
from our backward integration at time $t$. In general, for an arbitrary time 
$t$, the two determinations of $\Delta a_{j}$ from $\Delta {\Omega}$ and 
$\Delta {\varpi}$ will be different. As the time $t$ approaches the correct 
age of the family, the difference is expected to disappear. We use this 
method to determine the best estimate of $\tau$. Specifically, we define 
a ${\chi}^{2}$-like variable of the form
\begin{equation}
 {\chi}(t)=\sum_{j=2}^{N} \frac{\left|\Delta a_{j}^{\Omega}-\Delta a_{j}^{\varpi}
  \right|}{(N-1)}\; ,\label{eq: chi2}
\end{equation}
where $\Delta a_{j}^{\Omega}$ and $\Delta a_{j}^{\varpi}$ are the two 
determinations at time $t$, and search for the minimum of ${\chi}(t)$. When 
applied to the $N=576$ previously identified members of the Karin cluster, we 
found that the minimum occurs for $\tau = 5.746 \pm 0.011$ My. This result 
is in an excellent agreement with the age estimate of \citet{Nesvorny_2004} 
who found $\tau = 5.75 \pm 0.05$~My. The higher accuracy of our estimate 
is justified by the fact that our sample of the Karin cluster 
members is $\simeq$7 times larger than that of \citet{Nesvorny_2004}.

\begin{figure}
\epsscale{0.7}
\plotone{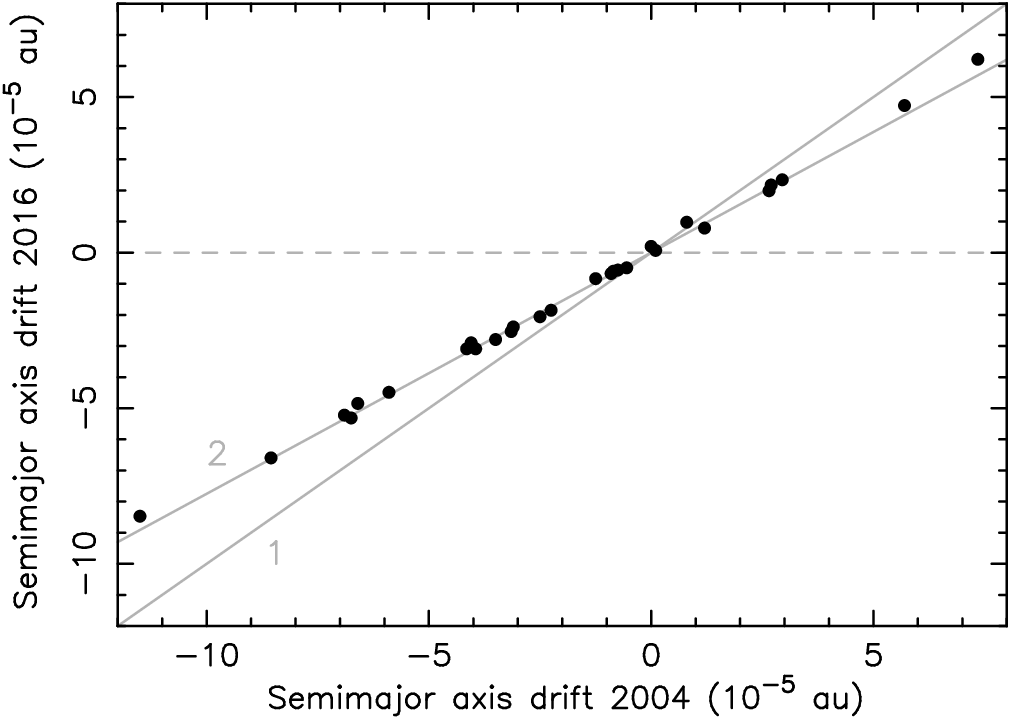}
\caption{Correlation between the drifts found in \citet{Nesvorny_2004}
 and those obtained from the present analysis. The gray line 1 has a slope
 $1$; the gray line 2 has a shallower slope $0.8$, implying the values obtained
 in this work are about $20$\% smaller (see the text for explanation).}
\label{fig: drift_correl}
\end{figure}

How well the semi-major axis drift rates determined here compare with those 
from \citet{Nesvorny_2004}? To answer this question we computed the mean value 
$\Delta a_{j}=(\Delta a_{j}^{\Omega}+\Delta a_{j}^{\varpi})/2$ for each individual 
member and compared these results with those obtained in \citet{Nesvorny_2004}.
Figure~\ref{fig: drift_correl} shows the result of this comparison. There is a 
very good correlation between the drift values obtained back in 2004 and here. 
Unfortunately, \citet{Nesvorny_2004} explicitly listed the $\Delta a_{j}$ 
values obtained for $t = -5.7$ and $-5.8$~My, but not for the time 
corresponding to the best age estimate. To use these estimates in 
Fig.~\ref{fig: drift_correl}, we have computed the mean of these values. 
Since the drift rates obtained for these times are systematically higher 
than the ones for $t=-5.75$ My, the mean is also
slightly higher. This explains why in Fig.~\ref{fig: drift_correl} 
the estimates for $t=-5.75$ My obtained in our work are systematically 
higher, by about $\simeq 20$\%, than the values inferred from the 2004 work.

\begin{figure}
\epsscale{0.7}
\plotone{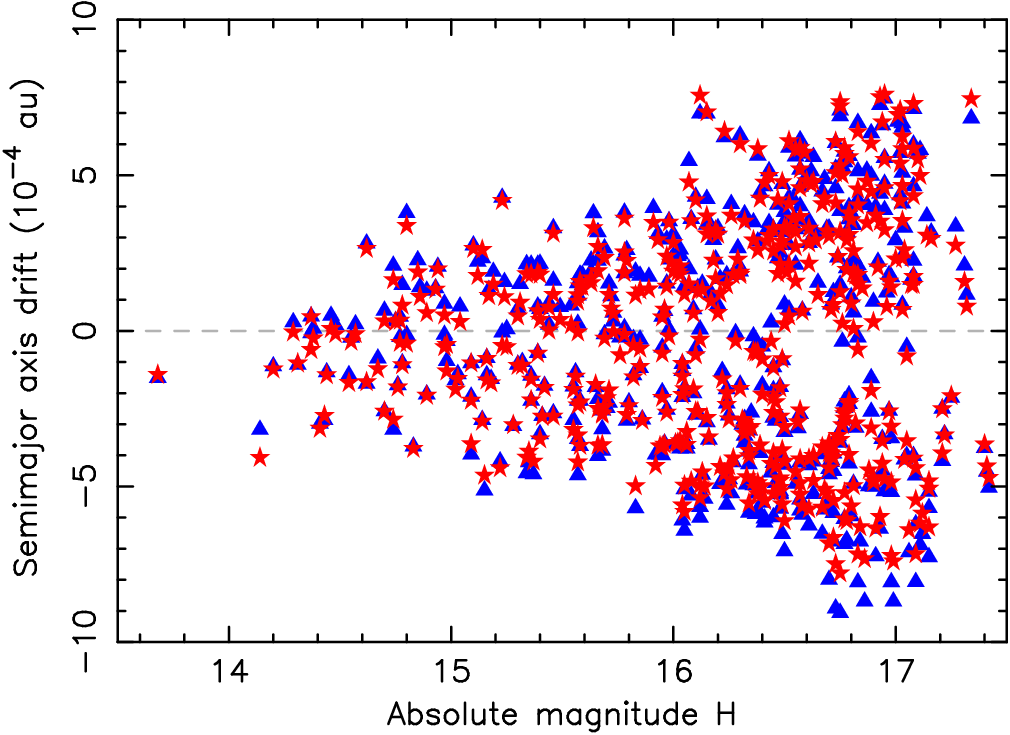}
\caption{Semi-major axis drift $\Delta a$ for 480 Karin cluster members
 that we inferred from the convergence of secular angles at $\tau = 5.746$ 
 My.  The blue triangles and the red stars denote the $\Delta a$ values 
 computed over the estimated family age from $\Delta \Omega$ and 
 $\Delta \varpi$, 
 respectively. There is a good consistency between the two determinations.}
\label{fig: drift_da}
\end{figure}

Figure~\ref{fig: drift_da} shows the $\Delta a_{j}$ values obtained here for 
the Karin cluster members, including hundreds of small members that were 
not known previously. As in \citet{Nesvorny_2004}, we eliminated from 
the original sample of 576 members all objects with orbital uncertainties 
in semi-major axis larger than $10^{-4}$ au, and those whose incompatibly 
large differences between $\Delta a_{j}^{\Omega}$ and $\Delta a_{j}^{\varpi}$ 
would suggest that they are probably interlopers (namely those with 
$\left|\Delta a_{j}^{\Omega}-\Delta a_{j}^{\varpi}\right| > 1.5 \times 10^{-4}$ au,
a value significantly larger than that for the vast majority of the studied
possible Karin member). The latter criterion eliminated only two objects.  

The results shown in Fig.~\ref{fig: drift_da} are in excellent agreement with 
Fig.~3 in \citet{Nesvorny_2004}. The measured magnitude of the semi-major 
axis drift increases with $H$, as expected for the Yarkovsky effect, 
whose strength is inversely proportional to the 
object diameter. The Yarkovsky drift magnitude over the estimated
age of the family nearly reaches $\simeq 10^{-3}$~au for the smallest 
members, which is just the right value for $D\simeq1$-2 km asteroids with 
extreme values of obliquities \citep[see below and][for more discussion]
{Bottke_2006}.

\section{The Bi-modality of Drift Rates} \label{sec: bim_rates}
The small members in Fig.~\ref{fig: drift_da} ($H\geq 16$-16.5) appear to 
have a bimodal distribution of the semi-major axis drifts with either 
relatively large positive or large negative values. This trend is 
reminiscent of the semi-major axis distribution found in several 
older asteroid families, where the distribution of the semi-major axis values 
is similarly bimodal. This trend has been interpreted as a result of 
the interplay between the Yarkovsky and YORP effects
\citep[see, e.g.,][]{Vokrouhlicky_2006a, Vokrouhlicky_2015, Nesvorny_2015}.
A similar line of reasoning suggests that Karin cluster is at 
the initial stage of this process.

Specifically, we suggest that the YORP effect acted on the Karin cluster 
members to slightly shift their obliquities toward extreme values 
($0^\circ$ and $180^\circ$), and this affected the overall magnitude of the 
accumulated Yarkovsky drifts. Obviously, the $\Delta a$ values measured in
Section 2 are relatively small ($<10^{-3}$~au; Fig.~\ref{fig: drift_da}), and 
the Yarkovsky effect has not altered the overall structure of the Karin 
family in proper element space.  Instead, the small change of the semi-major 
axes has only influenced the convergence of angles (as we discussed in the 
previous section). Before we present a detailed model of the Yarkovsky and 
YORP effects in Sect.~\ref{sec: Karin_yorp}, here we verify whether
the measured magnitude of drifts is in agreement with our theoretical 
expectations for the Yarkovsky effect.

\begin{figure}
\epsscale{0.7}
\plotone{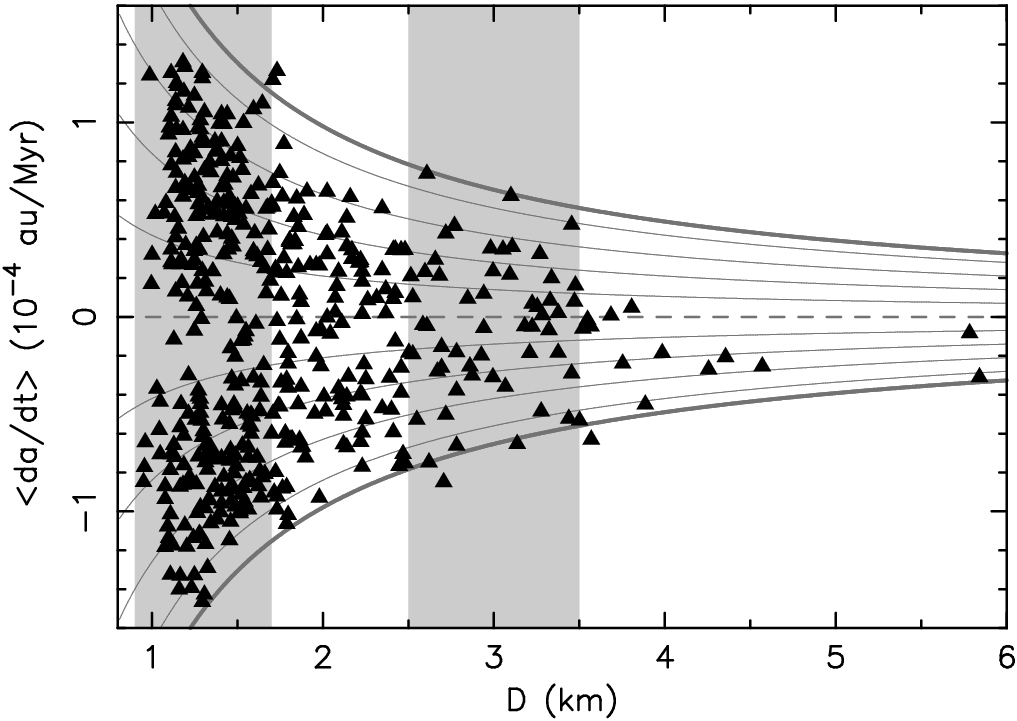}
\caption{The effective drift rate $\langle da/dt\rangle$ of
 the Karin cluster members (ordinate) vs their diameter $D$ (abscissa). 
 The gray lines are isolines of $\langle da/dt\rangle \propto 1/D$.
 If we choose $D=1.4$~km as reference value, the thin lines correspond to 
 $\langle da/dt\rangle$ values of $\pm (3,6,9,12)\times 10^{-5}$~au/My. 
 The thick gray lines, approximately enclosing all data-points, correspond 
 to $\langle da/dt \rangle = \pm 1.4\times 10^{-4}$~au/My for 
 $D=1.4$~km. The two size ranges, shown by the light gray rectangles, are 
 $D=0.9-1.7$~km (denoted I1) and $D=2.5-3.5$~km (denoted I2). The interval I1
 contains 280 data-points, while I2 contains 55 data-points. The distribution
 of drift rates in I1 is clearly bimodal with only a few bodies 
 with $\langle da/dt\rangle \simeq 0$. The drift rates in I2 are roughly 
 evenly distributed.}
\label{fig: dadt_d}
\end{figure}

First, in Fig.~\ref{fig: dadt_d}, we divide the accumulated drifts $\Delta a_j$ 
by the family age $\tau$, obtaining the effective drift rate 
$\langle da/dt\rangle_j = \Delta a_j/\tau$ for each Karin member.  A distinct 
characteristic of the Yarkovsky effect is that the drift rate is inversely 
proportional to body's diameter $D$. Therefore, in Fig.~\ref{fig: dadt_d}, we 
also plot isolines of $1/D=$const (gray lines).  The highlighted gray lines 
correspond to a drift value $\pm 1.4 \times 10^{-4}$~au My$^{-1}$ 
for a $D=1.4$-km body. These isolines approximately envelope the 
distribution of measured $\langle da/dt\rangle$.

This trend has been noticed previously \citep{Nesvorny_2004}, but here we 
also characterize the distribution for $D=1-2$ km asteroids which were not 
known in 2004. Presumably, the Karin members with the $\langle da/dt\rangle$ 
values close to the enveloping lines have an extreme value of the obliquity, 
because the Yarkovsky effect is maximized for $\varepsilon=0^\circ$ or 
$\varepsilon=180^\circ$.  Asteroids with the $\langle da/dt\rangle$ values inside 
the zone bracketed by the enveloping lines should have intermediate values 
of the obliquity. Various complications of this simple interpretation arise 
because the semi-major axis drift rate due to the Yarkovsky effect
depends on other parameters as well (such as, e.g., the asteroid rotation 
period). Bodies with the same obliquity value can thus drift at (slightly) 
different speeds (see the next section).

\section{Maximum Drift Rates} \label{sec: Max_drift_rates}
Here we compare the measured maximum drift rates 
($\langle da/dt\rangle\simeq1.4\times 10^{-4}$~au My$^{-1}$ for $D=1.4$~km) with 
the Yarkovsky effect theory developed in \citet{Vokrouhlicky_1999}
\citep[see also][]{Vokrouhlicky_2015}. Assuming a large body limit 
(i.e., penetration depth of the diurnal thermal wave much smaller than 
the body size) and keeping just the diurnal variant of the Yarkovsky effect, 
we have
\begin{equation}
 \frac{da}{dt} \simeq \frac{4\alpha}{9}\frac{\Phi}{n}
  \frac{\Theta}{1+\Theta+\frac{1}{2}\Theta^2 \rule{0pt}{2.3ex}}
  \cos\varepsilon\;, \label{eq: dadt_diu}
\end{equation}
where $\alpha=1-A$, with $A$ being the Bond albedo, $\Phi=(\pi D^2 F)/(4mc)$, 
$F\simeq 166.4$~W m$^{-2}$ is the solar radiation flux at the mean heliocentric 
distance of the Karin cluster, $m$ is the asteroid mass, $c$ is the velocity of 
light, and $n$ is the orbital frequency.

Note that $\Phi\propto 1/D$ which provides the aforementioned  
proportionality of the Yarkovsky effects with $1/D$. The thermal 
parameter $\Theta = \Gamma\sqrt{\omega}/(\epsilon\sigma T_\star^3)$
depends on the surface thermal inertia $\Gamma$, rotation
frequency $\omega$, and surface infrared emissivity $\epsilon$,
the Stefan-Boltzmann constant $\sigma$ and sub-solar temperature
$T_\star=[\alpha F/(\epsilon\sigma)]^{1/4}$. 

While we could use the thermal inertia $\Gamma$ as an independent parameter, we 
follow the tradition of the Yarkovsky effect studies and express it as
$\Gamma=\sqrt{K \rho_{\rm s} C}$, where $K$ is the surface thermal conductivity, 
$\rho_{\rm s}$ is the surface density and $C$ the surface thermal
capacity. For the sake of definiteness we fix $\rho_{\rm s}=2$~g cm$^{-3}$
and $C=680$~J/kg/K, and consider the thermal conductivity $K$
to be a free parameter (instead of $\Gamma$). The relationship
$da/dt \propto \cos\varepsilon$ gives the dependence of the Yarkovsky effect
on obliquity. Obviously, the maximum drift rates will occur for 
$\varepsilon=0^\circ$ (maximum positive rate) and $\varepsilon=180^\circ$
(maximum negative rate).

\begin{figure}
\epsscale{0.7}
\plotone{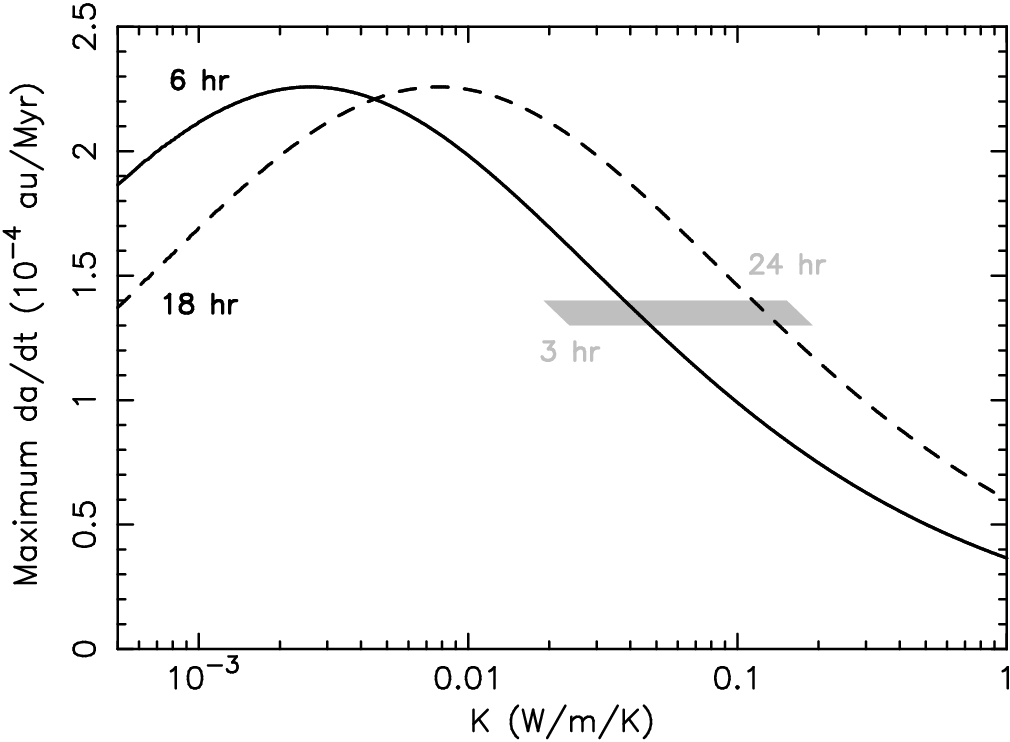}
\caption{Theoretical value of the diurnal Yarkovsky drift rate
 $da/dt$ at zero obliquity (Eq.~\ref{eq: dadt_diu}) as a
 function of the surface thermal conductivity for $D=1.4$~km. 
 We assumed Bond albedo $A=0.1$,
 thermal emissivity $\epsilon=0.9$, bulk density $\rho_{\rm b}
 =2.5$~g cm$^{-3}$, surface density $\rho_{\rm s}=2$~g cm$^{-3}$ and
 heat capacity $C=680$~J kg$^{-1}$K$^{-1}$. The rates were computed for two values
 of the rotation period, $P=6$~hr (solid line) and $P=18$~hr (dashed
 line). Because $da/dt$ is a function of $K/P$, the results can be easily  
 rescaled to other periods. The gray trapezoid highlights 
 $da/dt=(1.3-1.4)\times 10^{-4}$~au My$^{-1}$, which is roughly the range of 
 the maximum drift rates in Fig.~\ref{fig: dadt_d}.}
\label{fig: dadt_t}
\end{figure}

We now use Eq.~(\ref{eq: dadt_diu}) to compute the values $da/dt$
that would be expected for the $D=1.4$~km Karin members. For definiteness,
we assume $A=0.1$, $\epsilon=0.9$, and bulk density 
$\rho_{\rm b}=2.5$~g cm$^{-3}$. 
The rotation rate $\omega$ and thermal conductivity $K$ are varied within a 
reasonable range of values. The maximum drift rate of the Yarkovsky effect 
is obtained with $\varepsilon=0^\circ$. Figure~\ref{fig: dadt_t} shows the 
results.  To illustrate things we chose two typical values of the rotation
period: $6$~hr (solid line) and $18$~hr (dashed line). The gray trapezoid 
in Fig.~\ref{fig: dadt_t} is where the maximum drift rates are 
similar to the maximum drift rates inferred from small members of the Karin 
cluster ($(1.3-1.4)\times 10^{-4}$~au My$^{-1}$).

We note that the maximum $\langle da/dt\rangle$ values inferred from the small 
Karin cluster members are fully reasonable.  In fact, they are somewhat 
smaller than the optimal Yarkovsky drift rate for $D=1.4$~km Karin members 
that could be as large as $\simeq 2.2\times 10^{-4}$~au My$^{-1}$ (for low 
surface thermal inertia).  The measured values of $(1.3-1.4)\times 10^{-4}$~au 
My$^{-1}$ (Fig.~\ref{fig: dadt_d}) can be used to constrain the thermal 
conductivity/inertia. Assuming the typical rotation periods between 3 and 
24~hr, the measured value correspond the surface thermal conductivity in 
the range $0.02-0.2$~W m$^{-1}$K$^{-1}$ (Fig.~\ref{fig: dadt_t}).
This translates to the thermal inertia values 
$\simeq (170-500)$~J m$^{-2}$K$^{-1}$s$^{-1/2}$. 
These results are consistent with the determination of the thermal inertia 
for small near-Earth asteroids \citep[e.g.,][and M.~Delb\`o 
updates, personal communication]{Delbo_2007}.

\section{Prograde vs. Retrograde Rotators} \label{sec: rotators}
We now collect the $\langle da/dt\rangle$ measurements in the two 
highlighted size intervals shown in Fig.~\ref{fig: dadt_d}: (i) 
interval I1 with $D=0.9-1.7$~km, and (ii) interval I2 with $D=2.5-3.5$~km. 
The former contains 280 measurements, while the latter contains 55 
measurements. The primary data-set that we use here to analyze 
the YORP effect is I1. The set I2 is a control case that we use to make 
sure that our model (see below) consistently fits data for large sizes as 
well (note that I2 was roughly the size range available in 
\citet{Nesvorny_2004}). 

Figure~\ref{fig: data_i1i2} shows the distribution of $\Delta a$ values in the 
zones I1 (top) and I2 (bottom). In I1, there are 139 and 141 data points with 
negative and positive values of $\Delta a$, respectively. Recalling that this 
reflects the sign of $\cos\varepsilon$ (see Eq.~\ref{eq: dadt_diu}), we 
therefore find that an approximately equal number of small Karin cluster 
members has prograde and retrograde rotation.  This is interesting: the 
measurement of the drift rate for larger members indicates that there are 
more retrograde rotators among the largest fragments.  For example, the 
six members with $D>4$~km, except for (832)~Karin itself, are inferred to 
have a retrograde rotation \citep{Nesvorny_2004}.  (832)~Karin itself rotates 
in a prograde sense with a long rotation period \citep[e.g.,][]{Slivan_2012}.
This asymmetry, however, already disappears for the interval of sizes 
corresponding to I2, where there are 29 and 26 cases with negative and 
positive values of $\Delta a$, respectively. 

The median $\Delta a$ values for the negative and positive rotators in I1 are 
$\simeq -4.3\times 10^{-4}$~au and $\simeq 3.4 \times 10^{-4}$~au. 
Thus the peak of negative values is slightly more extended than the peak of 
positive values. There may be a physical reason for this.  Part of the
difference could be caused by the neglected drift of 832 Karin itself.
However, considering that the maximum drift of Karin computed
using \citet{Vokrouhlicky_1999} model of the Yarkovsky effect, the 
WISE estimated diameter, and the values of the parameters of the Yarkovsky
force from \citet{Broz_2013} is of the order of $6 \times 10^{-5}$~au, i.e.,
smaller than the observed difference, other mechanisms may be at play.
Recall that the 
obliquity evolution of the prograde rotators can be influenced by the 
spin-orbit resonances \citep[e.g.,][]{Vokrouhlicky_2003, Vokrouhlicky_2006b}.
If various other parameters such as the rotation period and dynamical 
ellipticity are favorable for capture in a resonance, the obliquity may 
end up oscillating around an equilibrium resonant point
\citep[e.g.,][]{Vokrouhlicky_2003}. This may halt the usual YORP-driven 
obliquity evolution of prograde rotators toward the extreme values, and 
produce an asymmetry of the accumulated drifts (note that the retrograde 
rotators are not subject to resonant capture; see Fig.~27 in 
\citet{Vokrouhlicky_2006b}). A detailed investigation of the spin-orbit 
dynamics is left for future work. 

\section{Comparison with Standard YORP Theory} \label{sec: YORP_theory}
Here we verify whether the YORP hypothesis for the origin of the bimodal 
distribution in the top panel of Fig.~\ref{fig: data_i1i2} is consistent 
with the standard YORP theory.  The strength of the YORP effect has a 
stronger dependence on $D$ than the Yarkovsky effect (it scales with 
$\propto 1/D^2$ rather than $\propto 1/D$ of the Yarkovsky effect;
e.g., \citet{Vokrouhlicky_2015}). This is why the Yarkovsky effect is detected 
in both size intervals I1 and I2, while the YORP-effect-induced bi-modality is 
apparent in I1 but not in I2. Assuming that the initial distribution of the 
spin vectors of small Karin members was isotropic, we estimate that the bimodal
distribution in I1 requires a characteristic change of $\simeq 0.5$ in
$\cos\varepsilon$ over the Karin cluster age. This roughly 
corresponds to an obliquity change of $\sim 30^\circ$-$40^\circ$. 

\citet{Capek_2004} modeled the YORP effect for a statistical sample of 
smooth Gaussian spheroids with $D=2$~km and a heliocentric distance 
$a=2.5$~au. Figure~11 in their paper shows that the maximum obliquity 
change of these bodies is typically $8.6^\circ$ per My (the maximum change 
happens for $\varepsilon\simeq35^\circ$).  An average rate for an
arbitrary obliquity is roughly one half of this value, or $4.3^\circ$ per 
My, which would accumulate to $\sim 25^\circ$ over the Karin family age.  
The YORP strength scales as $\propto 1/(D\,a)^2$.  Using this scaling we 
estimate that the obliquity of $D=1.4$~km asteroids (characteristic size 
in the interval I1) should have changed, on average, by $\sim 38^\circ$.
This is exactly what is required to explain the measured bi-modality in the 
interval I1. On the other hand, the estimated obliquity change of $D=3$~km 
bodies in the interval I2 is only $\sim 8^\circ$, which is clearly too 
small to appreciably affect the distribution.  

\section{The Yarkovsky-YORP Model} \label{sec: Karin_yorp}
Encouraged by the estimates discussed in the previous section, we now proceed 
by constructing a simple model for the Yarkovsky and YORP effects on small 
Karin cluster members.  We assume that the fragments initially created in the 
Karin-cluster formation event had: (i) an isotropic distribution of spin 
axis vectors, and (ii) their rotation rates were distributed according the 
Maxwellian distribution \citep[e.g.,][]{Pravec_2002}.  Impact simulations, 
such as the ones in \citet{Nesvorny_2006}, can be used to test whether (i) 
is reasonable. As for (ii), we note that the Maxwellian distribution
represents a good proxy for the distribution of rotation rates of 
fragments in the laboratory-scale impact experiments 
\citep[e.g.,][]{Giblin_1998}. 

In our simulations, we track the obliquity $\varepsilon$ and rotation-rate 
$\omega$ of each of the fragments as they evolve by the YORP effect.  The 
basic formulation of the YORP effect has been developed by 
\citet{Rubincam_2000}. \citet{Capek_2004} extended this approach to also 
include the effects of the surface thermal conductivity, and computed the 
characteristic YORP strength for a large sample of smooth irregular 
shapes (the so-called Gaussian spheroids).  Their results can be summarized 
as follows. 

The obliquity and rotation-rate evolution is given by two differential equations
\begin{eqnarray}
 \frac{d\omega}{dt} & = & f\left(\varepsilon\right) \; , 
  \label{eq: feq} \\
 \frac{d\varepsilon}{dt} & = & \frac{g\left(\varepsilon\right)}{\omega}
  \; ,  \label{eq: geq} 
\end{eqnarray}
where $f$ and $g$ are functions of obliquity. Each asteroid, having its own 
distinct shape, is described by different functional forms $f$ and $g$, but 
in a statistical sense the characteristic evolution can be obtained from 
the median functions derived in \v{C}apek \& Vokrouhlick\'y (2004).
In particular, we use the median values determined for the thermal 
conductivity $K=0.01$~W/m/K (see their Figs.~8 and 11 in \citet{Capek_2004}). 
In this setup, the obliquities always evolve toward the extreme values 
$\varepsilon=0^\circ$ and $180^\circ$, and the rotation rate may 
either increase or decrease when these asymptotic values are reached.  

\citet{Capek_2004} found that the tendency toward increasing or decreasing 
the rotation rate is roughly the same, at least for the statistical 
sample of asteroid shapes they tested.  This means that the value of 
the function $f$ is equally likely positive or negative when 
$\varepsilon=0^\circ$ or $180^\circ$. The $f$ and $g$ functions given in 
\citet{Capek_2004} are rescaled here to $D=1.4$~km (corresponding to I1) 
using $f\propto 1/D^2$ and $g\propto 1/D^2$.

Over the past decade a number of very detailed approaches have been developed 
to model the YORP effect \citep[see, e.g.,][for a review]{Vokrouhlicky_2015}.
One of the major findings of these works was a recognition that the small-scale 
surface irregularities can have an important contribution to the overall 
YORP strength. For example, the results of \citet{Rozitis_2012} and 
\citet{Golubov_2010} indicate that the $f$ and $g$ functions can have a 
somewhat smaller magnitude than the ones obtained for a smooth surface.
Additionally, a rough surface can trigger a tendency of the YORP effect to 
increase of the rotation rate. 

We introduce two empiric parameters in our YORP model to account for these 
complications (see \citet{Bottke_2015} for a similar approach).  First, we 
set $f=c_{\rm YORP}\,f_0$ and $g=c_{\rm YORP}\,g_0$, where $f_0$ and $g_0$ are the 
median functions from \citet{Capek_2004}, and $c_{\rm YORP}$ is a free 
strength parameter that expresses the actual strength of the YORP effect 
relative to $f_0$ and $g_0$. As noted above, we expect that $c_{\rm YORP}<1$. 
Second, we introduce an asymmetry parameter $\delta_{\rm YORP}$, defined as 
the fraction of bodies that undergo slow down of their rotation rate 
($1-\delta_{\rm YORP}$ is the fraction that is spun up).  The original model 
\citet{Capek_2004} gives $\delta_{\rm YORP}=0.5$, but considering the 
surface roughness, values $\delta_{\rm YORP}<0.5$ may be more appropriate.  
The best fit values of parameters $c_{\rm YORP}$ and $\delta_{\rm YORP}$ can be 
obtained from a fit to the measured distribution of obliquities.

We numerically integrate Eqs.~(\ref{eq: feq}) and (\ref{eq: geq}) using a 
simple Euler-type integration scheme with a time-step of $0.01$~My.  The 
initial obliquity and rotation period values are chosen on random.  Each 
simulation is repeated 10 times with different initial values.  The 
simulations are stopped at $\tau=5.746$~My, which is our best estimate of 
the Karin family age (Sec.~\ref{sec: karin_drift}).  As the time 
progresses, for each individual body we accumulate the change of the 
semi-major axis $\Delta a$ by the Yarkovsky effect from
\begin{equation}
 \Delta a_{\rm model} = \int_0^\tau \left(\frac{da}{dt}\right)\, dt = 
 \frac{4\alpha}{9}\frac{\Phi}{n} \int_0^\tau
  \frac{\Theta}{1+\Theta+\frac{1}{2}\Theta^2 \rule{0pt}{2.3ex}}
  \cos\varepsilon\, dt\;.  \label{eq: da_model}
\end{equation}
The parameters entering the right hand side of this equation were explained in
Sect. 5. Note that some of the variables, assumed to be constant, 
were pulled in front of the integral in (\ref{eq: da_model}), but some other
variables were left in the integrand (e.g., $\Theta$ and $\varepsilon$). 
Note that the latter parameters change due to the YORP effect. In particular, 
$\Theta\propto \sqrt{\omega}$.  To keep things simple, in each run we use a 
single value of the thermal surface conductivity $K$ for all bodies, but vary 
$K$ from one run to another. The bulk density of bodies is assumed to be 
$2.5$~g/cm$^3$. Below we will discuss how the results change for 
different density assumptions.

\begin{figure}
\epsscale{0.7}
\plotone{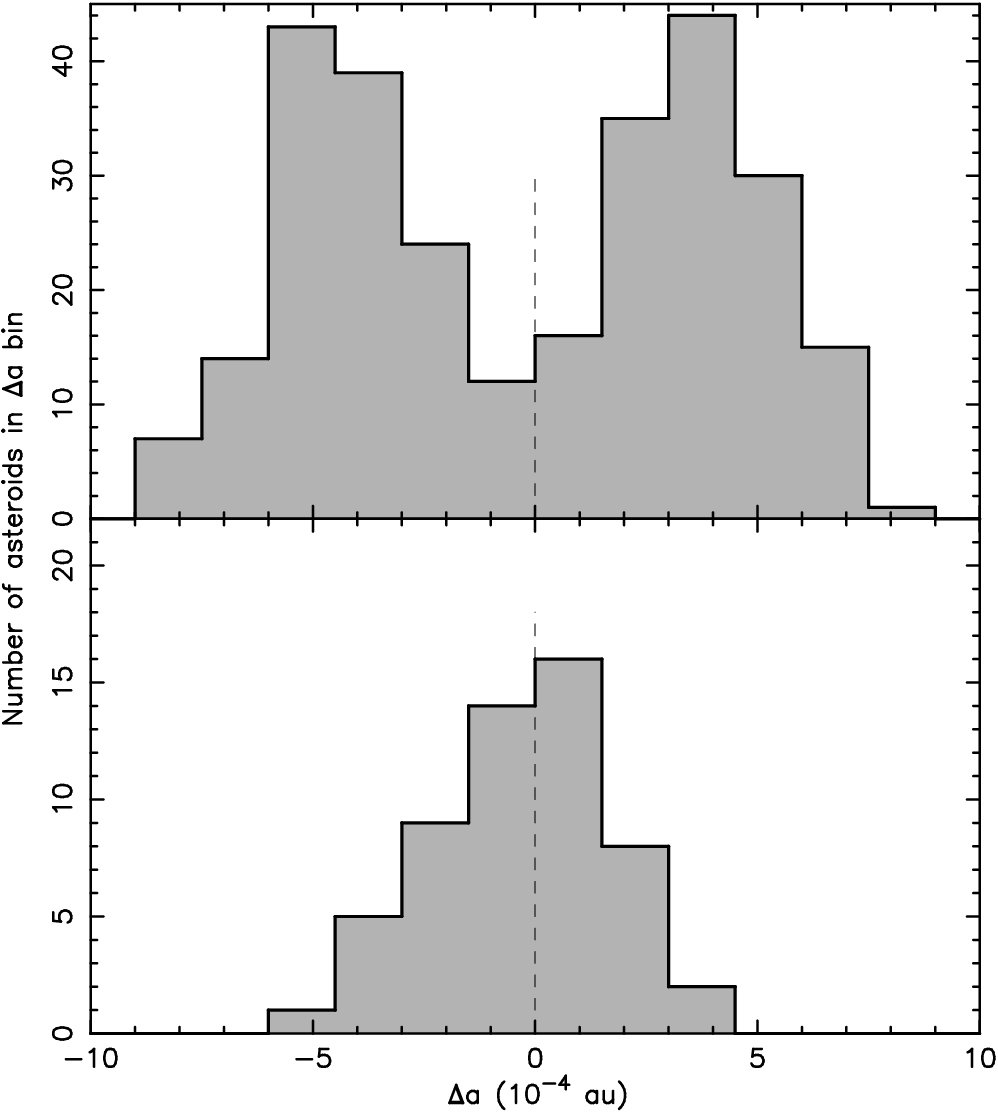}
\caption{The distribution of $\Delta a$ values in the intervals I1 (top)
 and I2 (bottom). Here we use a bin size of $1.5\times 10^{-4}$~au. Top:
 The sample contains 280 bodies with equally populated negative
 and positive values (139 vs. 141). The distribution is clearly bimodal.
 The median negative and positive values are $\simeq -4.3\times 10^{-4}$~au
 and $\simeq 3.4\times 10^{-4}$~au, respectively. Bottom: The sample 
 contains 55 bodies. There is no statistically significant difference 
 between the number of negative and positive values (29 vs
 26). Here the distribution is peaked at the origin.}
\label{fig: data_i1i2}
\end{figure}

Once the simulation is over, the model distribution of $\Delta a_{\rm model}$ 
values is compared with the measured distribution of $\Delta a$ shown in 
Fig.~\ref{fig: data_i1i2} (top panel). Because our model is not designed to 
reproduce any asymmetry in the distribution of obliquities (see discussion 
in Sec.~\ref{sec: karin_drift}), we modify the distribution of measured 
drifts by folding the negative and positive bins onto each other.  This 
leads to a symmetrical distribution shown by the red line in 
Fig.~\ref{fig: da_fit}. 

In each simulation, we fix $\delta_{\rm YORP}$ and run the model for different 
values of $c_{\rm YORP}$ and $K$ parameters.  We then attempt to minimize the 
difference between $\Delta a_{\rm model}$ and the measured $\Delta a$ 
distribution.  We use a bin size of $1.5\times 10^{-4}$~au (as in 
Fig.~\ref{fig: data_i1i2}), which leaves us with $N=12$ bins with useful 
information. Our minimization procedure uses a $\chi^2$-like target function:
\begin{equation}
 \chi^2 = \sum \left(\frac{n_{\rm model}-n}{\sigma_n}\right)^2\; ,
  \label{eq: tf}
\end{equation}
where the summation is performed over the 12 bins, $n$ is the number of
measurements and $n_{\rm model}$ the number of model bodies in each bin.

The denominator $\sigma_n$ expresses the uncertainty of each $n$ value.  A 
common practice is to set $\sigma_n\simeq\sqrt{n}$.  By adopting this 
assumption we find that our best-fit solutions would give $\chi^2 \simeq 13$, 
which is slightly larger than the number of bins.  This may mean that our 
simple model is incomplete or slightly inaccurate.  For example, as we 
discussed above, we do not model the effect of spin-orbit resonances
that may be important for the prograde rotators. It is also possible that a 
better result could be obtained if two $c_{\rm YORP}$ parameters were used, 
one that multiplies the $f$ function and one that multiplies the $g$ function.

Instead of investigating the possible physical reasons for this slight 
discrepancy, which would be a considerable work on its own, here we opted 
for a simple fix by setting $\sigma_n\simeq \sqrt{2n}$.  Our best fits 
give $\chi^2 \simeq 6.5$ with this definition. The confidence region in 
parameters $(K,c_{\rm YORP})$ around the best fit solution was 
defined as $\chi^2 < N$, where $N=12$.

\section{The Yarkovsky-YORP model: best-fitted parameters} \label{sec: results}
\begin{figure}
\epsscale{0.7}
\plotone{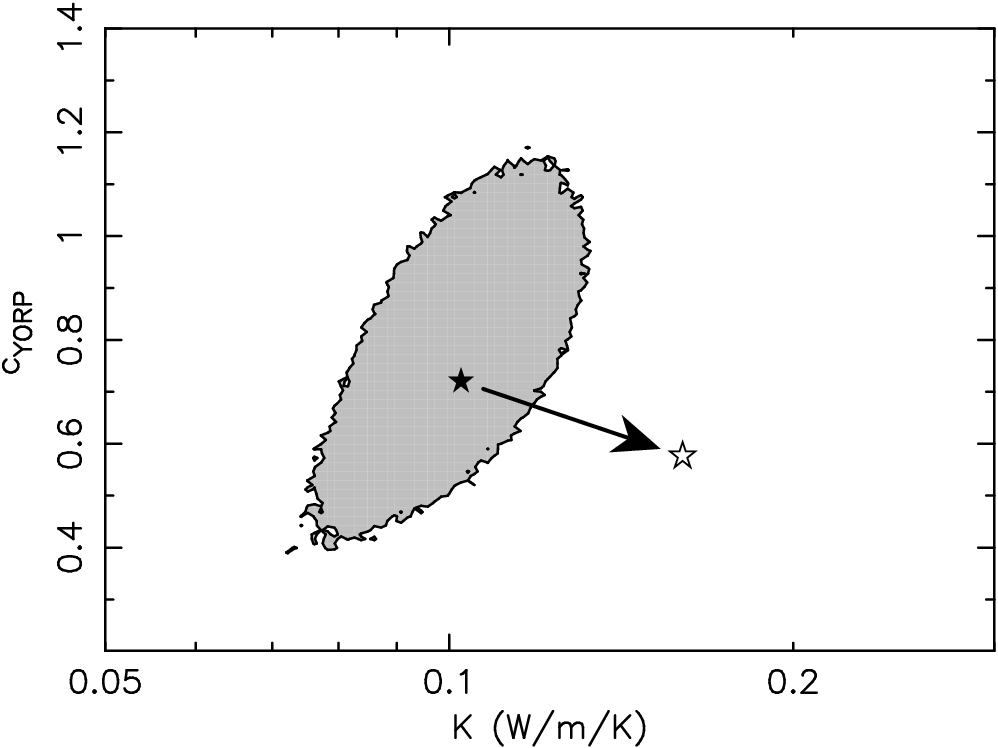}
\caption{The confidence interval defined as $\chi^2<N$ with $N=12$ (gray zone).
 The best fit solution is denoted by the black star. Here we fixed 
 $\delta_{\rm YORP}=0.4$ and varied the surface conductivity $K$ (abscissa) and 
 the $c_{\rm YORP}$ parameter. The bulk density was assumed to be $2.5$~g/cm$^3$. 
 If $\rho_{\rm b}=2$~g/cm$^3$ instead, the best fit solution would move as 
 indicated by the arrow, and the confidence region would shift as well.}
\label{fig: chi2_Kc}
\end{figure}

We found that the results only weakly depend on $\delta_{\rm YORP}$.  The best 
fits were obtained $0.3<\delta_{\rm YORP}<0.5$. We therefore fixed 
$\delta_{\rm YORP}=0.4$ in all subsequent simulations.  Figure~\ref{fig: chi2_Kc}
shows the main result of these simulations. The inferred values 
of the thermal conductivity range between $K=0.07$~W m$^{-1}$K$^{-1}$ and 
$0.13$~W m$^{-1}$K$^{-1}$, with the best-fit value $0.1$~W m$^{-1}$K$^{-1}$. 
Equivalently, the value of thermal inertia is found to be between $310$ and 
$420$~J m$^{-2}$K$^{-1}$s$^{-1/2}$. This range of values 
is consistent with (or perhaps only slightly larger than) the thermal inertia 
values estimated in \citet{Delbo_2007}. 

The confidence range of the $c_{\rm YORP}$ parameter is $0.4-1.1$, with the 
best-fit value of $0.72$. As discussed above such a value would be 
expected for a rough surface. It is also in a broad agreement with the 
results obtained for older asteroid families \citep[e.g.,][]{Vokrouhlicky_2006a}
and models of the pole and rotation rate 
distributions of small main-belt asteroids \citep[e.g.,][]{Hanus_2011}.

\begin{figure}
\epsscale{0.7}
\plotone{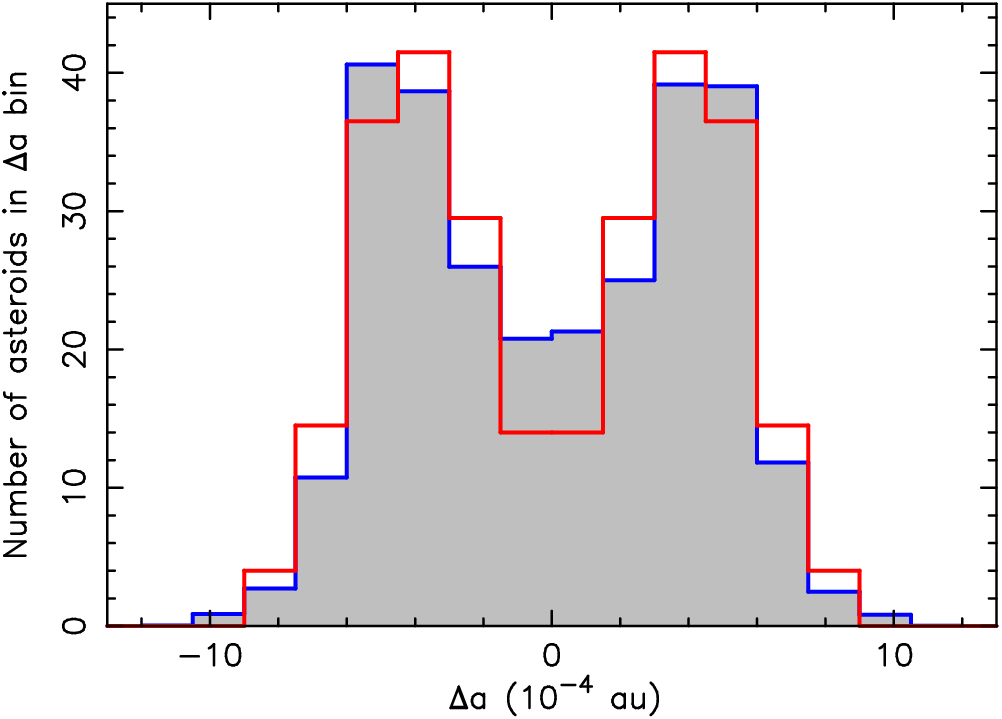}
\caption{The best fit solution for $\delta_{\rm YORP}=0.4$, $K^\star=0.1$ W m$^{-1}$
 K$^{-1}$ and $c_{\rm YORP}^\star=0.72$ (the gray histogram and blue line). The
 distribution of drift values inferred from the convergence criterion is shown
 by the red histogram.}
\label{fig: da_fit}
\end{figure}

Figure~\ref{fig: da_fit} shows how the best-fit solution compares with the
measured distribution of drift rates. The agreement is very good. A slight 
inconsistency arises in Fig. \ref{fig: da_fit} because the measured profile 
shows more depletion in the central bins with $\Delta a \simeq0$~au.  We suspect
that this points to a slight inconsistency of the assumed dependency of the
$g$ function on $\varepsilon$ for $\varepsilon \simeq 90^\circ$. Recall that
\citet{Capek_2004} computed the $g$ function for a specific collection of 
shapes.  It is possible that the young, freshly re-accumulated asteroids in 
the Karin family have a different distribution of shapes.  We leave this 
interesting problem for a future work.

\begin{figure}
\epsscale{0.7}
\plotone{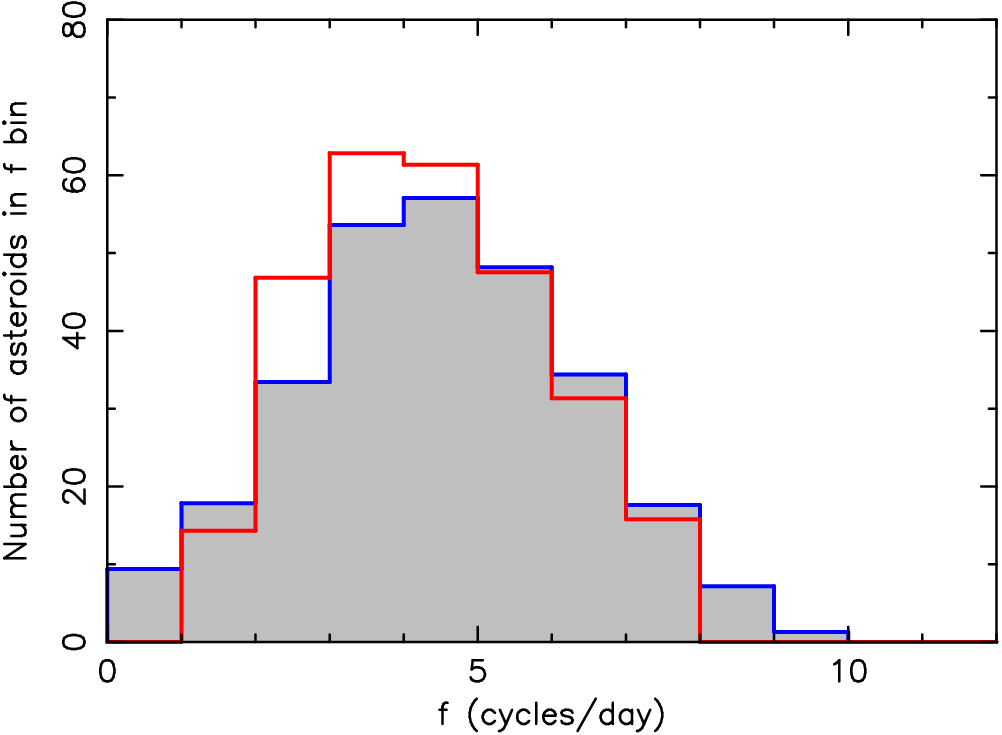}
\caption{Illustration of the YORP effect on the rotation
 frequencies in the best fit from Fig.~\ref{fig: da_fit}. The
 abscissa shows the rotation frequency $f=\omega/2\pi$ in cycles
 per day. The red histogram was the assumed initial distribution
 of the rotation frequencies. The gray histogram shows the final 
 distribution at $\tau=5.746$~My.}
\label{fig: dp}
\end{figure}

Figure~\ref{fig: dp} shows the distribution of the initial and final values of
the model rotation rates. We find that the small Karin members should still 
have roughly the same distribution of rotation rates as they had initially
just after the the family-formation event. 

Above we adopted the bulk density $\rho_{\rm b}=2.5$~g cm$^{-3}$.  In the 
subsequent simulations we tested the dependence of the results on 
$\rho_{\rm b}$ and found that the best-fit solution scales as 
$K^\star \propto 1/\rho_{\rm b}^2$ and $c_{\rm YORP}^\star\propto 
\rho_{\rm b}$ (the confidence regions recalibrate accordingly).  The scaling of 
$c_{\rm YORP}^\star$ arises from the YORP torque (inverse) dependence on 
body's mass.  The scaling of $K^\star$ is less transparent. It can be 
understood from the analysis of the Yarkovsky drift rate in semi-major axis 
given by Eq.~(\ref{eq: dadt_diu}).  Note that in the relevant regime of 
large $\Theta$ values, $da/dt\propto 1/(\rho_{\rm b} \Theta)$ and 
$\Theta\propto \sqrt{K}$.  Therefore, to have the same value of drift 
rate $da/dt$, $\rho_{\rm b}\sqrt{K}$ needs to be kept constant.  This produces 
the aforementioned scaling of the results. The arrow in 
Fig.~\ref{fig: chi2_Kc} indicates how the results would change if 
$\rho_{\rm b}=2$~g cm$^{-3}$.

Finally, we verified that our best-fit solution obtained for the size 
interval I1 does not violate constraints from the interval I2.  For that 
we used the best fit values of $(K,c_{\rm YORP},\delta_{\rm YORP})$, and re-run 
the simulations for $D=3$~km, which is a characteristic size in I2.  The 
modeled distribution of $\Delta a_{\rm model}$ was found to be consistent 
with the measured $\Delta a$ values shown in Fig.~\ref{fig: data_i1i2}
(bottom panel). We therefore confirm that the measured drift rates in I2 
were not significantly affected by the YORP effect.

\section{Numerical integration with the Yarkovsky effect and encounters 
 with Ceres} \label{sec: yarko_num}
The distribution of the semi-major axis drift rates were obtained in Sect.~3, 
where we used analytical arguments to improve the convergence of angles from 
a numerical simulation that ignored any drift. Here we include the semi-major 
axis drift directly in a numerical simulation to test how the convergence of 
angles is improved. In a separate simulation, we also include the gravitational 
effects of (1) Ceres to see how the convergence can be affected by close 
encounters of the Karin cluster members with Ceres.

We modified $SWIFT\_RMVS3$ \citep{Levison_1994} to include a semi-major 
axis drift \citep{Nesvorny_2004}.  For each of the confirmed Karin cluster 
members, we generated 13 clones with different drift rate values near the 
analytical estimate obtained in Sect.~\ref{sec: karin_drift}.  The orbits of 
the clones were tracked backward in time for 10 My.  We then checked which 
of the clones showed the best convergence of $\Omega$ and $\varpi$ at 
$\tau = 5.746$~My. The drift rate assigned to the best clones is our best 
numerical estimate of the actual drift rate. On one hand, the numerical rate 
inferred here can be considered a better estimate of the true drift rates
than the analytical method in Sect. 3.  In practice, however, the resolution 
with a limited number of clones is not good enough to distinguish between 
differences in the drift rates that are of the order of 5\%.  

\begin{figure}
\epsscale{0.8}
\plotone{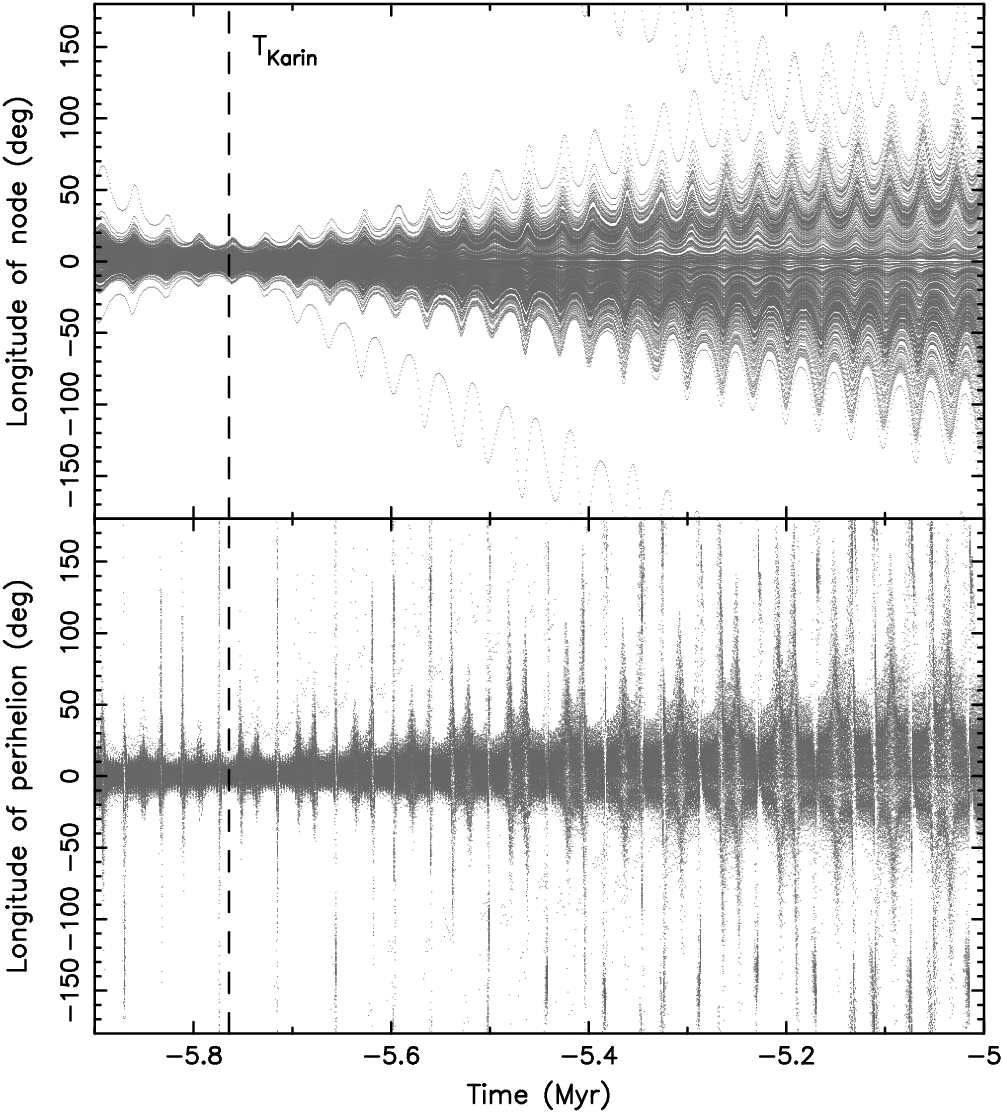}
\caption{The past orbital histories of 322 members of the Karin cluster:
 nodal longitude (top) and perihelion longitude (bottom). The values of 
 these angles are given here relative to (832) Karin.  Unlike in
 Fig.~\ref{fig: karin_angles}, here we accounted for the Yarkovsky effect
 explicitly in the integration. As a result, the convergence at $\tau=-5.764$ My
 has significantly improved.}
\label{fig: Karin_Yarko_angles}
\end{figure}

About 70\% of the best clones converged to within $\pm 10^{\circ}$ in 
$\Omega$ and $\varpi$ at $\tau = 5.746$~My. Their past orbital histories 
are shown in Fig.~\ref{fig: Karin_Yarko_angles}.  The remaining 30\% of the 
best clones converged as well, but not within $\pm 10^{\circ}$.  This is 
contributed by the limited resolution of our numerical integration and/or, 
at least in some cases, by short-period oscillations of the osculating 
angles $\Omega$ and $\varpi$ near the estimated family age. A more 
detailed study of this problem is left for future work. 

Next we discuss the results obtained when (1) Ceres was included in the 
numerical integration. There are two ways that (1) Ceres can be 
influencing the results.  First, a close encounter between a small 
asteroid and Ceres can lead to a change of the small body's semi-major axis, 
which would then influence the measured drift rate. Second, the secular 
resonances with (1) Ceres \citep{Novakovic_2015} can alter the precession 
rates of $\Omega$ and $\varpi$, and therefore influence the convergence of 
these angles as well.  To determine which of these effects has a bigger 
weight, we monitored in the simulation all close encounters of all bodies 
with (1) Ceres. 

\begin{figure}
\epsscale{0.7}
\plotone{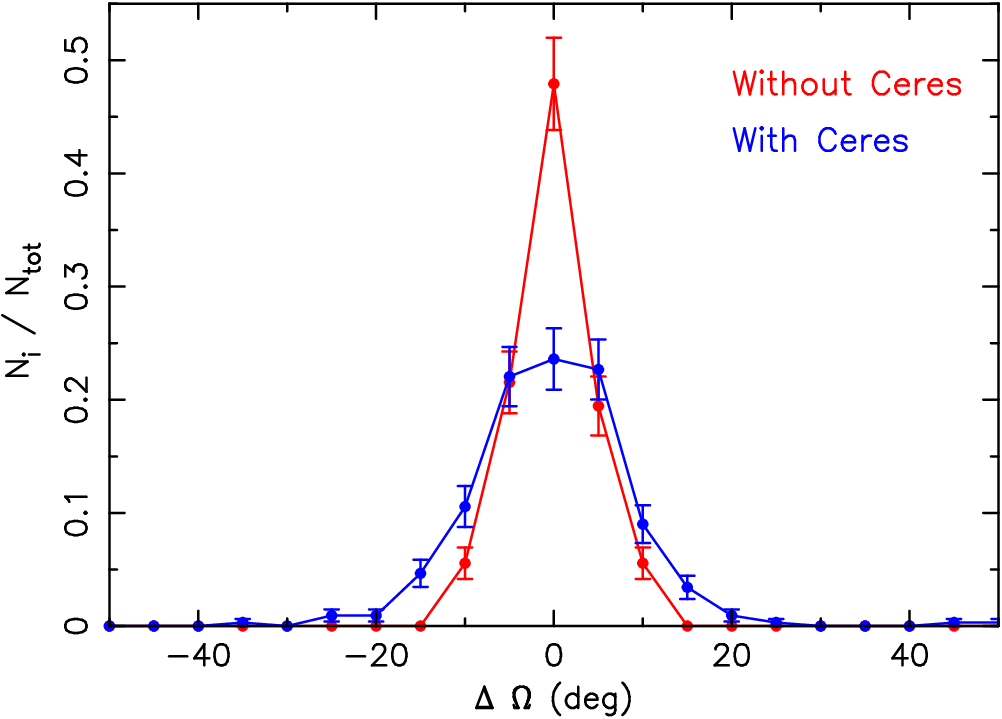}
\caption{The distribution of $\Omega$ values at $\tau=-5.764$ My for the
 cases without Ceres (red line) and with Ceres (blue line). The error bars are 
 assumed to be proportional to the square root of the number of objects in 
 each bin.} 
\label{fig: ceres_enc}
\end{figure}

Figure~\ref{fig: ceres_enc} compares the distribution of $\Omega$ values 
obtained for $\tau = 5.746$~My in our numerical simulations with and 
without Ceres (results for $\varpi$ are similar).  The distribution in the 
simulation with Ceres is clearly broader.  We find that this is mainly a 
consequence of close encounters with Ceres.  Of the 322 clones that converged 
within $\pm 10^\circ$ in a simulation without Ceres, roughly 80\% converge 
within $\pm 10^\circ$ in a simulation with Ceres.  The remaining 20\% of the 
best clones do not converge so well.  Of these, roughly 75\% suffered 
close encounters to Ceres (within the Hill sphere or closer).  A small 
fraction of clones suffered a very close Ceres encounter, and had 
$\Delta \Omega \simeq 30^\circ$ at $\tau = 5.746$~My.  On average, 
Ceres encounters add $\simeq 4^\circ$ to the dispersion of angles at 
the time of convergence. This limits the precision to which the convergence 
of angles can be determined, and therefore the accuracy with which
the $\Delta a$ values over the estimated age of the family can be computed: 
a difference of $4^\circ$ corresponds to a difference of $7.2 \times 10^{-5}$~au 
in the $\Delta a$ computed from the convergence of $\Omega$ and of 
$5.3  \times 10^{-5}$~au for that from $\varpi$.  Including other massive 
asteroids in the simulation would slightly increase this threshold.

\section{Conclusions} \label{sec: conc}
The main results of this work can be summarized as follows:
\begin{itemize}
\item We revised the Karin family membership using the identification 
from \citet{Nesvorny_2015} and initially including asteroids in the 
immediate neighborhood of the Karin cluster. The taxonomical and albedo 
interlopers were eliminated. We numerically integrated the orbits of all 
selected objects backward in time over 10 My. Using the convergence 
criteria described in the main text, we identified 576 asteroids that 
are very likely true members of the Karin cluster.

\item Using the method of Nesvorn\'y \& Bottke (2004), we inferred the 
drift rates caused by the Yarkovsky effect. By minimizing the difference 
between two determinations of $\Delta a$ from ${\Omega}$ and ${\varpi}$ 
we found that the age of the Karin cluster is $\tau=5.746 \pm 0.011$ My.  
This age determination is consistent with and improves on the previous 
estimate. 

\item Since the Yarkovsky drift rate depends on obliquity, we interpreted 
the observed distribution of the drift rates in terms of the obliquity 
distribution. For small, $D=1$-2 km Karin cluster members, the distribution 
of obliquities is clearly bimodal. The best explanation for such a 
distribution is that the YORP effect acted to alter the distribution that 
has been more uniform initially. 

\item We simulated the evolution of obliquities and spin rates with a simple
Yarkovsky/YORP model. We found that the magnitude of the obliquity changes 
required to explain the bimodal distribution is consistent with the YORP 
effect and inferred age $\tau$. The surface thermal conductivity is inferred to
 be $0.07-0.2$ W m$^{-1}$ K$^{-1}$, corresponding to the thermal inertia of 
$\simeq 300-500$ J m$^{-2}$ K$^{-1}$ s$^{-1/2}$). We find that the strength of the 
YORP effect is roughly $\simeq 0.7$ of the nominal strength obtained for a 
collection of random Gaussian spheroids. These results are consistent with a 
surface composed of rough, rocky regolith. 

\item We performed additional numerical simulations with the Yarkovsky drift 
and gravity of (1) Ceres. We found that the close encounters of Karin cluster
members with Ceres act to increase the dispersion of angles and does not allow
us, even in principle, to obtain a perfect convergence. On average, Ceres 
increases the dispersion of angles by $\sim 4^\circ$  at $\tau=5.746$~My. 
\end{itemize}

Our work motivates new observational efforts. In particular, it would be 
interesting to verify the obliquity distribution inferred from our work.  
A decade ago such a goal would have been a remote possibility, but recent 
advancements in asteroid shape and rotation state studies can lead to 
interesting results soon. For example, the obliquities of individual 
bodies can be obtained from the sparse photometry data of ground-based 
survey programs \citep[e.g.,][]{Durech_2016}. Even more powerful 
results are expected from the space missions such as Gaia 
\citep[e.g.,][]{Mignard_2007}.

\section*{Acknowledgments}
We thank the reviewer of this paper, Dr. Bojan Novakovi\'{c}, for 
comments and suggestions that improved the quality of this work.
This paper was written while the first author was a visiting scientist
at the Southwest Research Institute (SwRI) in Boulder, CO, USA.
We would like to thank the S\~{a}o Paulo State Science Foundation 
(FAPESP) that supported this work via the grant 14/24071-7.
D.N.'s work on this project was supported by NASA's Solar System Workings
program, while D.V. was supported by the Czech Grant Agency 
(grant GA13-01308S). This publication makes use of data products from 
the Wide-field Infrared Survey Explorer (WISE) and from NEOWISE, which are 
a joint project of the University of California, Los Angeles, and the 
Jet Propulsion Laboratory/California Institute of Technology, funded by 
the National Aeronautics and Space Administration.  

\section{Appendix 1} \label{sec: appendix_1}
\begin{figure}
\epsscale{0.7}
\plotone{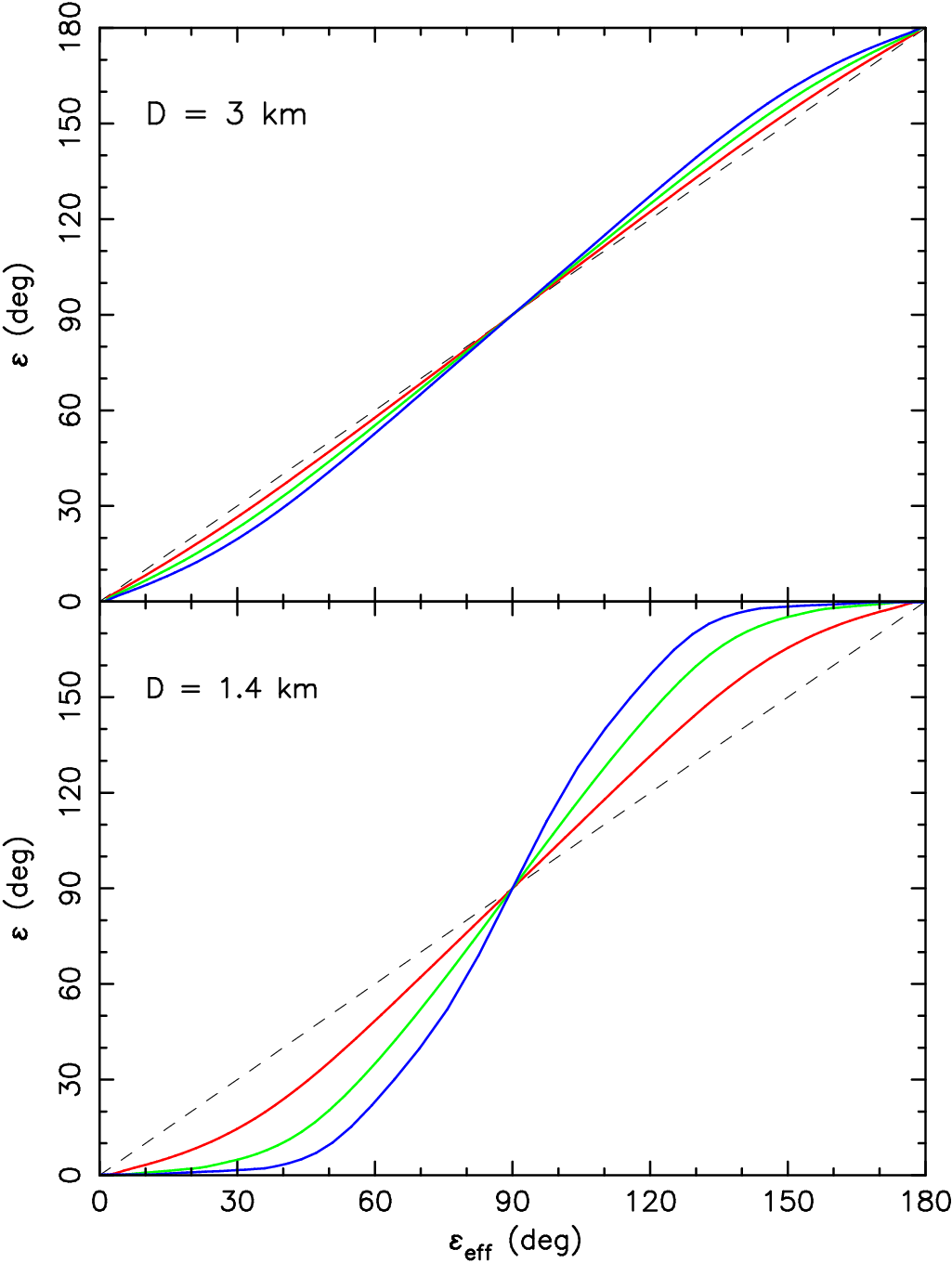}
\caption{Relation between the effective obliquity $\varepsilon_{\rm eff}$
 (column~9 of Table~1) and the current obliquity  $\varepsilon$ (ordinate) of 
 the simulated bodies in the Karin family.  The top panel is for $D=3$~km 
 members, a representative size in the interval I2, while the bottom panel 
 is for $D=1.4$~km members (representative for I1).  The three curves in 
 each of the panels were obtained for different rotation periods: $4$~hr 
 (red), $8$~hr (green), and  $12$~hr (blue). For a larger asteroid size in 
 the top panel, both obliquity values nearly coincide. For a smaller size, 
 the effective obliquity $\varepsilon_{\rm eff}$ can be slightly larger 
 (if $\varepsilon_{\rm eff}<90^\circ$) or smaller (if 
 $\varepsilon_{\rm eff}>90^\circ$) than the current value of 
 $\varepsilon$, especially if the rotation rate is slow.}
\label{fig: de}
\end{figure}

In Table~1 we report the list of 480 identified Karin 
cluster members. The table lists their absolute magnitude, proper elements 
$(a_P, e_P, \sin i_P)$, frequencies $g$ and $s$, Lyapunov exponents (Lyapunov
exponents with values near {\bf to $1.5 \times 10^{-6} yr^{-1}$} correspond 
to objects for which the integration time was not long enough to obtain a 
convergence), estimated mean obliquity $\varepsilon$, and estimated mean 
Yarkovsky drift speed.  Note to observers: the obliquity values listed in 
Table 1 are historical mean obliquities and may not exactly coincide 
with the current values. Figure~\ref{fig: de} shows the correspondence between 
the historical obliquity values given in Table~1 and our estimate of
the present obliquities.

\onecolumn
\begin{center}
\begin{longtable}{|c|c|c|c|c|c|c|c|c|c|}
\caption{Karin cluster members: absolute magnitudes, proper elements and 
frequencies, Lyapunov exponents, and estimated mean obliquities and
Yarkovsky drift speed.}
\label{table: karin_members}\\
\hline
Number & $H$ & $a_P$ & $e_P$ & $\sin i_P$ & $g_P$ & $s_P$  & $LCE$ & $\epsilon$ &  Drift speed \\
       &     & [au]  &       &           & $''$ yr$^{-1}$&$''$ yr$^{-1}$& $10^{-6}$ yr$^{-1}$&  [deg]  & $[10^{-5}$ au{My}$^{-1}]$ \\
\hline
\endfirsthead
\multicolumn{10}{|c|}%
{\tablename\ \thetable\ -- \textit{Continued from previous page}} \\
\hline
Number & $H$ & $a_P$ & $e_P$ & $\sin i_P$ & $g_P$ & $s_P$  & $LCE$ & $\epsilon$ &  Drift speed \\
       &     & [au]  &       &           & $''$ yr$^{-1}$&$''$ yr$^{-1}$& $10^{-6}$ yr$^{-1}$&  [deg]  &  $[10^{-5}$ au{My}$^{-1}]$ \\
\hline
\endhead

\hline \multicolumn{10}{|r|}{{Continued on next page}} \\ \hline
\endfoot

\hline 
\endlastfoot
   832 &  11.18 &    2.86440 &    0.04390 &    0.03687 &   70.639 &  -65.235 &   1.49 &     &         \\
 10783 &  13.41 &    2.86480 &    0.04411 &    0.03683 &   70.676 &  -65.270 &   1.49 &  113 &   -0.84 \\
 11728 &  13.48 &    2.86555 &    0.04443 &    0.03676 &   70.746 &  -65.334 &   1.56 &  180 &   -3.10 \\
 13765 &  14.41 &    2.86967 &    0.04585 &    0.03688 &   71.112 &  -65.667 &   1.15 &  180 &   -5.33 \\
 13807 &  13.68 &    2.86885 &    0.04527 &    0.03682 &   71.037 &  -65.590 &   1.92 &  162 &   -2.54 \\
 15649 &  14.70 &    2.86418 &    0.04393 &    0.03650 &   70.632 &  -65.226 &   1.47 &  180 &   -4.50 \\
 16706 &  14.43 &    2.86200 &    0.04330 &    0.03676 &   70.435 &  -65.051 &   1.49 &  180 &   -4.86 \\
 20089 &  14.76 &    2.86170 &    0.04312 &    0.03696 &   70.402 &  -65.022 &   1.46 &  139 &   -3.10 \\
\hline
\end{longtable}
\end{center}
%\twocolumn

\clearpage

\clearpage

\clearpage

\clearpage

\clearpage

\clearpage

\clearpage

\clearpage

\clearpage

\clearpage

\clearpage

\clearpage

\clearpage


\begin{thebibliography}{}

\bibitem[Bottke et al.(2002)]{Bottke_2002} Bottke, W. F., Vokrouhlick\'{y}, 
 D., Rubincam, D.P., Bro\v{z}, M. 2002, in Asteroids III, ed. W. F. 
Bottke et~al. (Tucson: University of Arizona Press), 395

\bibitem[Bottke et al.(2006)]{Bottke_2006} Bottke, W.~F.,
 Vokrouhlick{\'y}, D., Rubincam, D.~P., \& Nesvorn{\'y}, D.\ 2006, Annu. Rev. 
 Earth Planet. Sci., 34, 157

\bibitem[Bottke et al.(2015)]{Bottke_2015} Bottke, W. F., Vokrouhlick\'y, D.,
 Walsh, K. J., et~al. 2015, Icarus, 247, 191

\bibitem[Bro\v{z}(1999)]{Broz_1999} Bro\v{z}, M. 1999, Master Thesis,
 Charles University, Prague, Czech Republic
 
\bibitem[Bro\v{z} et al.(2013)]{Broz_2013} Bro\v{z}, M., Morbidelli, A., 
 Bottke, W.~F., et~al. 2013, A\&A, 551, A117

\bibitem[Bus \& Binzel(2002a)]{Bus_2002a} Bus, S.~J., \& Binzel, R.~P.\ 2002a,
 Icarus, 158, 106

\bibitem[Bus \& Binzel(2002b)]{Bus_2002b} Bus, S.~J., \& Binzel, R.~P.\ 2002b,
 Icarus, 158, 146

\bibitem[\v{C}apek \& Vokrouhlick\'{y}(2004)]{Capek_2004} \v{C}apek, D., 
 \& Vokrouhlick\'y, D. 2004, Icarus, 172, 526

\bibitem[Carruba et al.(2013)]{Carruba_2013} Carruba, V. Huaman, M. E., 
 Domingos, R. C., \& Roig, F. 2013, A\&A, 550, A85

\bibitem[Carruba et al.(2015)]{Carruba_2015} Carruba, V., Nesvorn\'{y}, D., 
 Aljbaae, S., \& Huaman, M. E. 2015, MNRAS, 451, 4763

\bibitem[Delb\`o et al.(2007)]{Delbo_2007} Delb\`o, M., Dell'Oro, A., 
 Harris, A. W., Mottola, S., \& Mueller, M. 2007, Icarus, 190, 236

\bibitem[DeMeo \& Carry(2013)]{DeMeo_2013} DeMeo, F.~E., \& Carry, 
 B.\ 2013, Icarus, 226, 723

\bibitem[\v{D}urech et al.(2016)]{Durech_2016} \v{D}urech, J., 
 Hanu\v{s}, J., Oszkiewicz, D., \& Van\v{c}o, R. 2016, A{\&}A, 587, A48

\bibitem[Giblin et al.(1998)]{Giblin_1998} Giblin, I., Martelli, G.,
 Farinella, P., et~al. 1998, Icarus, 134, 77

\bibitem[Golubov et al.(2010)]{Golubov_2010} Golubov, O., \& Krugly, Y. N.
 2010, ApJLett, 752, L11.

\bibitem[Hanu\v{s} et al.(2011)]{Hanus_2011} Hanu\v{s}, J., \v{D}urech,
 J., Bro\v{z}, M., et~al. 2011, A{\&}A, 530, 134

\bibitem[Harris et al.(2009)]{Harris_2009} Harris, A.~W., Mueller, 
 M., Lisse, C.~M., \& Cheng, A.~F.\ 2009, Icarus, 199, 86

\bibitem[Ivezi{\'c} et al.(2001)]{Ivezic_2001} Ivezi{\'c}, {\v Z}., 
 Tabachnik, S., Rafikov, R., et~al. 2001, AJ, 122, 2749

\bibitem[Lazzaro et al.(2004)]{Lazzaro_2004} Lazzaro, D., Angeli, 
 C.~A., Carvano, J.~M., et~al. 2004, Icarus, 172, 179

\bibitem[Levison \& Duncan(1994)]{Levison_1994} Levison, H. F., \& Duncan, 
 M. J. 1994, Icarus, 108, 18

\bibitem[Masiero et al.(2012)]{Masiero_2012} Masiero, J.~R., Mainzer, A.~K.,
 Grav, T., Bauer, J.~M., \& Jedicke, R.\ 2012, ApJ, 759, 14

\bibitem[Mignard et al.(2007)]{Mignard_2007} Mignard, F., Cellino, A.,
 Muinonen, K., et~al. 2007, EM{\&}P, 101, 97

\bibitem[Milani \& Kne\v{z}evi\'{c}(1994)]{Milani_1994} Milani, A., \&
 Kne\v{z}evi\'{c}, Z. 1994, Icarus, 107, 219

\bibitem[Molnar \& Haegert(2009)]{Molnar_2009} Molnar, L.~A., \&
 Haegert, M.~J.\ 2009. AAS/Division for Planetary Sciences Meeting Abstracts 
 \#41 41, \#27.05 

\bibitem[Murray \& Dermott(1999)]{Murray_1999} Murray, C.~D., \&
 Dermott, S.~F.\ 1999, Solar System Dynamics (Cambridge University Press,
 Cambridge)

\bibitem[Nesvorn{\'y} et al.(2002)]{Nesvorny_2002} Nesvorn{\'y}, D., 
 Bottke, W.~F., Dones, L., \& Levison, H.~F.\ 2002, Nature, 417, 720

\bibitem[Nesvorn\'{y} \& Bottke(2004)]{Nesvorny_2004} Nesvorn\'{y}, D., \&
 Bottke, W. F. 2004, Icarus, 170, 324
 
\bibitem[Nesvorn{\'y} et al.(2006)]{Nesvorny_2006} Nesvorn{\'y}, D., Enke,
 B.~L., Bottke, W.~F., et~al. 2006, Icarus, 183, 296 

\bibitem[Nesvorn\'y et al.(2015)]{Nesvorny_2015} Nesvorn\'{y}, D., Bro\v{z}, 
 M., \& Carruba, V.\ 2015, in Asteroids IV, ed. P. Michel et~al. (Tucson:
 University of Arizona Press), 297

\bibitem[Novakovi{\'c} et al.(2012)]{Novakovic_2012} Novakovi{\'c}, B., Hsieh, 
H.~H., \& Cellino, A.\ 2012, \mnras, 424, 1432 

\bibitem[Novakovi\'{c} et al.(2015)]{Novakovic_2015} Novakovi\'{c}, B., 
 Clara, M., Tsirvoulis, G., \& Kne\v{z}evi\'{c}, Z. 2015, ApJ, 807, L5

\bibitem[Pravec et al.(2002)]{Pravec_2002} Pravec, P., Harris, A. W., 
 \& Michalowski, T. 2002, in Asteroids~III, ed. W. F. Bottke et~al. 
 (Tucson: University of Arizona Press), 113

\bibitem[Pravec et al.(2008)]{Pravec_2008} Pravec, P., Harris, A. W.,
 Vokrouhlick\'y, D., et~al. 2008, Icarus, 197, 497

\bibitem[Press et al.(2001)]{Press_2001} Press, V. H., Teukolsky, S. A.,
 Vetterlink, W. T., \& Flannery, B. P. 2001, Numerical Recipes in
 Fortran 77 (Cambridge University Press, Cambridge)

\bibitem[Rayman(2015)]{Rayman_2015} Rayman, M. D., Dawn Journal, May 28,
 2015

\bibitem[Rozitis \& Green(2012)]{Rozitis_2012} Rozitis, B., \& Green, S. F.
 2012, MNRAS, 423, 367

\bibitem[Rubincam (2000)]{Rubincam_2000} Rubincam, D. P. 2000, Icarus, 
 148, 2

\bibitem[\v{S}idlichovsk\'{y} \& Nesvorn\'{y}(1997)]{Sidli_1997} 
 \v{S}idlichovsk\'{y}, \& M., Nesvorn\'{y}, D. 1997, Cel. Mech. Dynam.
 Astron., 65, 137

\bibitem[Slivan \& Molnar(2012)]{Slivan_2012} Slivan, S. M., \& Molnar, L. A.
 2012, Icarus, 220, 1097

\bibitem[Tholen(1989)]{Tholen_1989} Tholen, D.~J.\ 1989, in Asteroids II, 
 ed. R. P. Binzel et~al. (Tucson: University of Arizona Press), 1139

\bibitem[Vokrouhlick\'{y}(1999)]{Vokrouhlicky_1999} Vokrouhlick\'{y}, D.
 1999, A\&A, 334, 362

\bibitem[Vokrouhlick\'{y} et~al.(2003)]{Vokrouhlicky_2003} Vokrouhlick\'y, D.,
 Nesvorn\'y, D., \& Bottke, W. F. 2003, Nature, 425, 147

\bibitem[Vokrouhlick\'{y} et~al.(2006a)]{Vokrouhlicky_2006a} Vokrouhlick\'y, D., 
 Bro\v{z}, M., Bottke, W. F., Nesvorn\'y, D., \& Morbidelli, A. 2006a, 
 Icarus, 182, 118

\bibitem[Vokrouhlick\'{y} et~al.(2006b)]{Vokrouhlicky_2006b} Vokrouhlick\'y, D., 
 Nesvorn\'y, D., \& Bottke, W. F. 2006b, Icarus, 184, 1

\bibitem[Vokrouhlick\'{y} et~al.(2015)]{Vokrouhlicky_2015} Vokrouhlick\'y, D., 
 Bottke, W. F., Chesley, S. R., Scheeres, D. J., \& Statler, T. S. 2015,
 in Asteroids~IV, ed. P. Michel et~al. (Tucson: University of Arizona Press),
 509

\bibitem[Xu et al.(1995)]{Xu_1995} Xu, S., Binzel, R.~P., Burbine, T.~H.,
 \& Bus, S.~J.\ 1995, Icarus, 115, 1

\bibitem[Zellner et al.(1985)]{Zellner_1985} Zellner, B., Tholen, 
 D.~J., \& Tedesco, E.~F.\ 1985, Icarus, 61, 355
    
\end{thebibliography}
\end{document}